\documentclass[aps,prd,superscriptaddress,twoside,twocolumn,nofootinbib,%
10pt,showpacs,floatfix]{revtex4-1}

\usepackage{amsmath,amssymb}
\usepackage{graphicx,bm}
\usepackage{ulem}
\usepackage{slashed}
\usepackage{dcolumn}
\usepackage{epsfig}
\usepackage{multirow}

\allowdisplaybreaks

\begin{document}


\title{Tetraquark mixing framework for isoscalar resonances in light mesons}


\author{Hungchong Kim}%
\email{hungchong@kau.ac.kr}
\affiliation{Research Institute of Basic Science, Korea Aerospace University, Goyang, 412-791, Korea}

\author{K. S. Kim}
\affiliation{School of Liberal Arts and Science, Korea Aerospace University, Goyang, 412-791, Korea}

\author{Myung-Ki Cheoun}%
\affiliation{Department of Physics, Soongsil University, Seoul 156-743, Korea}

\author{Makoto Oka}%
\affiliation{Department of Physics, Tokyo Institute of Technology, Meguro 152-8551, Japan }
\affiliation{Advanced Science Research Center, Japan Atomic Energy Agency, Tokai, Ibaraki, 319-1195 Japan}

\date{\today}


\begin{abstract}

Recently, a tetraquark mixing framework has been proposed for light mesons and applied more or less successfully to
the isovector resonances,
$a_0(980), a_0(1450)$, as well as to the isodoublet resonances, $K^*_0(800), K^*_0(1430)$.
In this work, we present a more extensive view on the mixing framework
and apply this framework to the isoscalar resonances, $f_0 (500)$, $f_0(980)$, $f_0 (1370)$, $f_0(1500)$.
Tetraquarks in this framework can have two spin configurations
containing either spin-0 diquark or spin-1 diquark and each configuration forms a nonet in flavor space.
The two spin configurations are found to mix strongly through the color-spin interactions.
Their mixtures,
which diagonalize the hyperfine masses, can generate the physical resonances constituting two nonets,
which, in fact, coincide roughly with the experimental observation.
We identify that $f_0 (500)$, $f_0(980)$ are the isoscalar members
in the light nonet, and $f_0 (1370)$, $f_0(1500)$ are the similar members in the heavy nonet.
This means that the spin configuration mixing, as it relates the corresponding members in the two nonets, can generate $f_0 (500), f_0 (1370)$
among the members in light mass, and $f_0(980), f_0(1500)$ in heavy mass.
The complication arises because the isoscalar members of each nonet are subject to an additional flavor mixing known as
OZI rule
so that $f_0 (500), f_0 (980)$, and similarly $f_0 (1370), f_0 (1500)$, are
the mixture of two isoscalar members belonging to an octet and a singlet in SU$_f$(3). The tetraquark mixing framework
including the flavor mixing is tested for the isoscalar resonances in terms of the mass splitting and the fall-apart decay modes.
The mass splitting among the isoscalar resonances is found to be consistent qualitatively with their hyperfine mass splitting
strongly driven by the spin configuration mixing, which suggests that the tetraquark mixing framework works.
The fall-apart modes from our tetraquarks also seem to be consistent with the experimental modes.  We also discuss
possible existence of the spin-1 tetraquarks that can be constructed by the spin-1 diquark.

\end{abstract}


\maketitle

\section{Introduction}

Multiquarks, normally referred to hadrons composed by four or higher number of quarks,
are a subject of the intensive study recently because the
newly discovered resonances especially in the heavy quark sector might be the candidates for them.
This interest has been triggered by the observation of $X(3872)$~\cite{Belle03,Aubert:2004zr, Choi:2011fc, Aaij:2013zoa}
which, among various possibilities, could be the tetraquarks with the flavor structure
$cq\bar{c}\bar{q}~ (q=u,d)$~\cite{Maiani:2004vq,Kim:2016tys}.
For the other resonances newly measured,
$X(3823)$, $X(3900)$, $X(3940)$, $X(4140)$, $X(4274)$, $X(4500)$, $X(4700)$,
reported in Ref.~\cite{Bhardwaj:2013rmw,Xiao:2013iha,Abe:2007jna, Aaij:2016iza,Aaij:2016nsc},
one promising scenario would be tetraquarks also.
In addition, pentaquarks are an interesting topic in the multiquark study.
In particular, $P_c(4380)$, $P_c(4450)$ are recently observed as bumps in the $J/\psi~ p$ channel from the $\Lambda_b^0 \rightarrow J/\psi K^- p$
decay~\cite{Aaij:2015tga}
and they are the strong candidates for pentaquarks because their quark content, guessing from the decay mode, is expected to be $uudc\bar{c}$.

Although multiquarks candidates are accumulating among those resonances recently
discovered in the heavy quark sector,
they may not be necessarily limited to new resonances with heavy quarks.
The reason why the multiquark candidates are more common for the new resonances with heavy quarks
is mainly due to their peculiar decay modes that allow such an interpretation more clearly.
In principle, they also need to be found in the existing hadrons from those normally regarded as excited
states because, after all, multiquarks are composite objects of quarks
bounded by color forces that are basically independent of quark flavors.

A long-standing example is the nonet
consisted of $a_0 (980)$, $f_0 (500)$, $f_0 (980)$, $K_0^* (800)$
in the light meson system
which are regarded as the strong candidates for tetraquarks with the diquark-antidiquark form~\cite{Jaffe77a,Jaffe77b,Jaffe04,Jaffe:1999ze,MPPR04a,EFG09}.
In addition, a tetraquark picture for the excited states of $D$, $B$ mesons
seems to work fine with some intriguing description of the mass splitting as well as their decay patterns~\cite{Kim:2014ywa}.
Ultimately, it would be nice if one can come up with a unified framework for
multiquarks that can be applied not only to light mesons but also to heavy mesons.

The diquark-antidiquark model is most popular for tetraquarks and
this could be the best candidate for a unified framework in the end.
However, to solidify this framework further even in the light meson system,
there are some issues to be clarified related to possible diquarks especially in comparison with the meson spectroscopy.
In the original construction of this model, the diquark, belonging to spin-0, color antitriplet, flavor antitriplet,
is used to construct a tetraquark nonet.
As is well known, the strong candidates for this picture are the lowest-lying resonances in the $J^{P}=0^{+}$ channel,
namely, $a_0 (980)$, $f_0 (500)$, $f_0 (980)$, $K_0^* (800)$.

This picture can be updated by the following observation.
In Particle Data Group (PDG)~\cite{PDG16},
one can find another resonances with the same spin-parity but with higher masses,
$a_0 (1450)$, $f_0 (1370)$, $f_0 (1500)$, $K_0^* (1430)$, which can be regarded as an another nonet.
Later, we will see that this nonet satisfies certain characteristics of Jaffe's tetraquarks.
In our point of view, it is quite tempting to combine these
resonances into a tetraquark framework also.
One possible way is
to construct additional tetraquarks using the less compact diquark with spin-1, color sextet, flavor antitriplet.
These tetraquarks also form a nonet in flavor and, in principle, they can be matched to the heavy nonet here.

But, an important aspect that we want to address is that two tetraquarks may not be
the eignestates of the Hamiltonian.
Specifically, they can mix through color-spin interactions which then generate
the off-diagonal hyperfine masses.
The physical states, therefore, must be identified as the eigenstates that diagonalize the hyperfine masses.
In other words, the two nonets in the meson spectroscopy are linear combinations of the two tetraquarks.
This tetraquark mixing framework seems to work for the isovector and isodoublet resonances,
$a_0 (980), a_0 (1450)$, $K_0^* (800), K_0^* (1430)$. Their hyperfine mass splitting
matches relatively well with their experimental mass splitting~\cite{Kim:2016dfq} and
the fall-apart modes of $a_0 (980), a_0 (1450)$ seem to be consistent with
their experimental decay modes~\cite{Kim:2017yur}.

In this work, we test this tetraquark mixing framework further by applying it to the isoscalar resonances.
The difficulty in this extension is the additional flavor mixing between the octet and singlet members
of SU$_f$(3), normally known as Okubo-Zweig-Iizuka (OZI) rule.
Below, we consider the flavor mixing issue in three different cases.
The first one is to assume the exact SU$_f$(3) symmetry where there is no flavor mixing.
The second is the ideal mixing case where the flavor mixing occurs maximally so that the strange quarks
are completely decoupled from the nonstrange quarks in the tetraquark systems.
The flavor mixing parameters are fixed uniquely in the ideal mixing case.
The third one is the general flavor mixing where the mixing parameters will be fitted to the
mass splitting of the physical resonances.

Our tetraquark mixing framework is quite different from
other approaches that can be found in the literature for the resonances of our concern.
To name a few, Refs.~\cite{Boglione:2002vv,Wolkanowski:2015lsa,vanBeveren:1986ea,Dudek:2016cru} investigated
the isovector resonances, $a_0(980)$ and $a_0(1450)$, as the pole structures generated dynamically
by a single ${\bar q}q$ state or
from coupled-channel meson-meson scattering.
Ref.~\cite{Maiani:2006rq} considered the heavy nonet above as the tetraquarks mixed with a glueball
while Ref.~\cite{Giacosa:2006tf} treated $a_0 (980)$ as mixtures of tetraquarks and quarkonia.
In Ref.~\cite{Black:1999yz}, the $a_0(1450)$, $K_0^*(1430)$ are treated by the $P$-wave $\bar{q}q$ mixed with the four-quark
$qq\bar{q}\bar{q}$ scalar nonet.
The alternative tetraquark approach including instantons has been proposed in
Refs.~\cite{Dorokhov:1989hm,Dorokhov:1993nw} where four
possible tetraquarks are considered including the $27_f$ flavor multiplet.
Judging from various approaches, the current status is rather unclear on the nature of the resonances
being considered here.
Nevertheless, we believe that our tetraquark framework provides a relatively simple picture
that can be tested easily in terms of reproducing the mass splitting and decay modes of the resonances of concern.

This paper is organized as follows.
In Sec.~\ref{sec:motivation}, we motivate the tetraquark mixing framework based on the meson spectra listed in PDG.
We introduce two possible tetraquarks in the diquark-antidiquark form utilizing the two diquark
configurations.
The spin-1 diquark configuration necessarily requires additional tetraquarks to be found
in spin-1 and spin-2 channels.
We discuss in Sec.~\ref{sec:spin12} the possible candidates for the spin-1,2 tetraquarks in PDG.
Then after introducing the hyperfine mass and its connection to the mass formula in Sec.~\ref{sec:mass},
we test our mixing framework in generating the mass splitting for the spin-0 tetraquarks in Sec.~\ref{sec:spin0}.
Sec.~\ref{sec:hf_spin12} presents the hyperfine masses for spin-1 and spin-2 tetraquarks.
In Sec.~\ref{sec:fall apart}, we provide fall-apart decay modes of our tetraquarks as a further testing ground of our mixing framework.
We summarize in Sec.~\ref{sec:summary}.

\section{Motivation for tetraquark mixing framework}
\label{sec:motivation}

In this section, we revisit the tetraquark mixing framework advocated in Ref.~\cite{Kim:2016dfq} but in a wider
perspective.  The presentation here is more extensive in a sense that we are considering
the full tetraquark nonet in motivating
the mixing framework while the discussion in Ref.~\cite{Kim:2016dfq} was limited to
isovector and isodoublet resonances. This will eventually help in understanding how
tetraquarks are realized in hadron spectroscopy.

First, we start by examining briefly the tetraquark model of
Jaffe~\cite{Jaffe77a,Jaffe77b,Jaffe04,Jaffe:1999ze} in which the spin-0 tetraquarks are constructed by
combining diquarks ($q q$) and antidiquarks ($\bar{q}\bar{q}$).
In this construction, the diquark is in a state with spin-0, color antitriplet ($\bar{\bm{3}}_c$), and flavor antitriplet ($\bar{\bm{3}}_f$)
because the diquark with this type, which we call the spin-0 diquark, is most compact among all the
possible diquarks.
This can be inferred from the binding energies of the diquarks calculated from the color-spin interactions~\cite{Jaffe:1999ze}.
The resulting tetraquarks, in a diquark-antidiquark form, $q q \bar{q}\bar{q}$, have the spin configuration
\begin{equation}
|J,J_{12},J_{34}\rangle=|000\rangle,
\end{equation}
where $J$ is the spin of the tetraquark, $J_{12}$ the diquark spin, $J_{34}$ the antidiquark spin.
Note that all the quarks are assumed to be in an $S$-wave state in this approach
because tetraquarks being considered are supposed to be in the ground state.
The color configuration is $|\bm{1}_c,\bar{\bm{3}}_c,\bm{3}_c\rangle$,
which can be written in terms of the individual quark color as
\begin{eqnarray}
\frac{1}{\sqrt{12}}  \varepsilon_{abd}^{} \ \varepsilon^{aef}
\Big ( q^b q^d \Big )
\Big ( \bar{q}_e \bar{q}_f \Big )\ ,
\label{color wave function1}
\end{eqnarray}
where the roman indices denote the colors.

\begin{figure}[t]
\centering
\epsfig{file=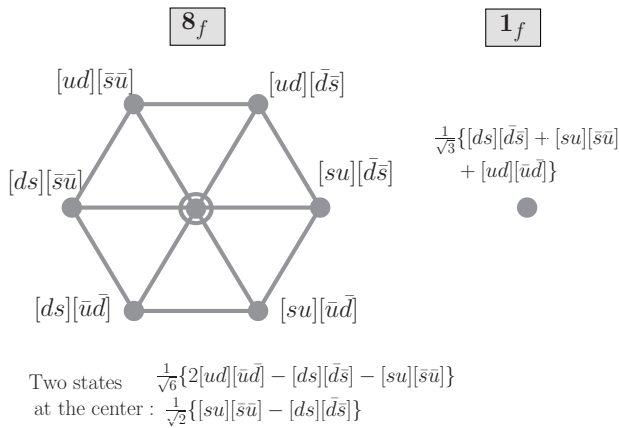, width=0.95\hsize}
\caption{Flavor structure of the tetraquark nonet. }
\label{flavor structure}
\end{figure}

The tetraquarks in flavor space form a nonet, $\bar{\bm{3}}_f\otimes \bm{3}_f=\bm{8}_f\oplus \bm{1}_f$.
By adopting a tensor notation for the
flavor multiplets~\footnote{Ref.~\cite{Oh:2004gz} might be useful for technical details in using the tensor notation.},
their members can be expressed as
\begin{eqnarray}
[{\bf 8}_f]^i_{j} &=& T_{j}\bar{T}^{i}-\frac{1}{3} \delta^{i}_{j}~ T_{m}\bar{T}^{m}\label{flavor octet}\ ,\\
{\bf 1}_f &=& \frac{1}{\sqrt{3}}T_{m}\bar{T}^{m}\label{flavor singlet}\ .
\end{eqnarray}
Here the diquark ($T_i$) [the antidiquark ($\bar{T}^i$)] is an antisymmetric combination of
quarks [antiquarks] given by
\begin{eqnarray}
T_i &=&\frac{1}{\sqrt{2}}\epsilon_{ijk}q_j q_k\equiv [q_j q_k]\ ,\nonumber\\
\bar{T}^i &=& \frac{1}{\sqrt{2}}\epsilon^{ijk}\bar{q}^j \bar{q}^k \equiv [{\bar q}^j {\bar q}^k] \ .
\end{eqnarray}
Fig.~\ref{flavor structure} shows a weight diagram of the nonet with explicit quark flavors.
So the wave functions of the tetraquarks in this approach are completely determined in terms of spin, color, flavor space.

Assuming that all the quarks are in an
$S$-wave state, the tetraquarks have the following characteristics.
\begin{enumerate}
\item All the members in the nonet have spin-0 with positive parity, $J^{P}=0^{+}$, by their construction.
\item Being nonet members, the possible isospins are $I=0,1/2,1$.
\item By counting the number of strange quarks in the wave functions, it is expected that the isovector ($I=1$) members
are heavier than the isodoublet ($I=1/2$) members. For example, $[su][{\bar d} {\bar s}]$ is expected to be heavier than $[su][{\bar u} {\bar d}]$.
\item Using the flavor wave functions given in Fig.~\ref{flavor structure}, one can determine $C$-parity for the members
with $I_z =0$. As one can see in Appendix, it can be proven to be positive, i.e. $J^{PC}=0^{++}$ for the $I_z=0$ members.
\end{enumerate}
Thus, corresponding candidates in the actual spectrum must be sought from
the resonances satisfying these constraints.
In fact, the lowest-lying states with $J^P=0^+$ in PDG as collected in the top part of Table~\ref{spin0++},
$a_0 (980)$, $f_0 (500)$, $f_0 (980)$, $K_0^* (800)$, seem to form
a nonet satisfying the four characteristics described above.  They have the expected quantum numbers such as isospin, spin-parity, and $C$-parity.
More importantly, the isovector members are heavier than the isodoublet members, $M(a_0) > M(K_0^*)$, coinciding with the third characteristics above.
We stress that this type of mass ordering can not be established from a simple two-quark system, $q\bar{q}$.
Therefore, $a_0 (980)$, $K_0^* (800)$, $f_0 (500)$, $f_0 (980)$, are
the strong candidates for the tetraquark nonet~\cite{Jaffe77a,Jaffe77b,Jaffe04,Jaffe:1999ze}.


\begin{table}
\centering
\begin{tabular}{c|c|c|c|c}  \hline\hline
 $J^{PC}$ & $I$ &  Meson  & Mass(MeV) & $\Gamma$(MeV) \\
\hline
\multicolumn{5}{c}{[Lowest-lying resonances]} \\
\hline
\multirow{3}{*}{$0^{++}$}
         & $ 0 $ & $f_0 (500)$  & 400-550 &  400-700   \\
         & $ 0 $ & $f_0 (980)$  & 990 &  10-100  \\
         & $ 1 $ & $a_0 (980)$  & 980 &  50-100   \\
\hline
$0^{+}$  & $1/2$ & $K_0^* (800)$ & 682 & 547  \\
\hline
\multicolumn{5}{c}{[Higher resonances]} \\
\hline
\multirow{8}{*}{$0^{++}$}
         & $ 0 $ & \underline{$f_0 (1370)$} & 1200-1500 &  200-500   \\
         & $ 1 $ & \underline{$a_0 (1450)$} & 1474 &  265  \\
         & $ 0 $ & \underline{$f_0 (1500)$} & 1505 &  109   \\
         & $ 0 $ & $f_0 (1710)$ & 1723 &  139   \\
         & $ 1 $ & $a_0 (1950)$ & 1931 &  271 \\
         & $ 0 $ & $f_0 (2020)$ & 1992 &  442   \\
         & $ 0 $ & $f_0 (2100)$ & 2101 &  224   \\
         & $ 0 $ & $f_0 (2200)$ & 2189 &  238   \\
         & $ 0 $ & $f_0 (2330)$ & 2314 &  144   \\
\hline
\multirow{2}{*}{$0^{+}$}
        & $1/2$ & \underline{$K_0^* (1430)$} & 1425 & 270  \\
        & $1/2$ & $K_0^* (1950)$ & 1945 & 201  \\
\hline\hline
\end{tabular}
\caption{Resonances of $J^{PC}=0^{+}$ collected from PDG~\cite{PDG16}. The upper part is the lowest-lying resonances
which are the strong candidates for the tetraquark nonet.  The bottom part is higher resonances and
the underlined members among them
can be selected as additional candidates for the tetraquark nonet.}
\label{spin0++}
\end{table}

But what we want to point out is that this nonet in the lowest-lying states is not the only possibility.
In fact, there are additional candidates
for the tetraquark nonet with higher masses in PDG. As one can see from the bottom part of Table~\ref{spin0++},
there are various resonances with higher masses in $J^P=0^+$. Among them,
the resonances with relatively lower masses,
$a_0 (1450)$, $K_0^* (1430)$, $f_0 (1370)$, $f_0 (1500)$, seem
to form an another nonet satisfying the four characteristics described above.
Specifically, these resonances have the anticipated quantum numbers, such as isospin, spin-parity, and $C$-parity.
The isovector member $a_0 (1450)$, although marginal, is still heavier than the isodoublet member $K_0^* (1430)$ by 50 MeV.
Therefore, $a_0 (1450)$, $K_0^* (1430)$, $f_0 (1370)$, $f_0 (1500)$ might be the 2nd candidates for the tetraquark nonet.
This selection among higher resonances seems to be unique because these are well separated in mass from the rest
resonances.
In passing, it may be worth mentioning that these members are much heavier than the nonet members in
the lowest-lying states, $a_0 (980)$, $K_0^* (800)$, $f_0 (500)$, $f_0 (980)$, by more than 500 MeV or so.
So there are huge mass gaps between the two nonets.

Now, we have two nonets in PDG that satisfy the tetraquark characteristics.
Therefore, it is quite tempting to combine the two nonets in a tetraquark framework.
If one attempts to do so, one can immediately see that the spin-0 diquark alone, even though it is the
optimal building block in constructing tetraquarks, is not enough to explain the two nonets.
We need an additional building block for tetraquarks.
For this purpose, it may be possible to use the spin-1 diquark
with the color and flavor structure, ($\bm{6}_c$,$\bar{\bm{3}}_f$).
This spin-1 diquark is the second most compact object among four possible diquarks~\cite{Jaffe:1999ze}.
Furthermore, since its flavor structure is $\bar{\bm{3}}_f$, tetraquarks constructed from this spin-1 diquark also
form a nonet similarly as the heavy nonet in PDG.

Then, we can use the two diquarks
\begin{eqnarray}
&&\text{Spin-0~diquark}:~ J=0, \bar{\bm{3}}_c, \bar{\bm{3}}_f\label{spin0diquark}\ ,\\
&&\text{Spin-1~diquark}:~ J=1, \bm{6}_c, \bar{\bm{3}}_f \label{spin1diquark}\ ,
\end{eqnarray}
in constructing tetraquarks.
These two diquarks share a common fact that their binding energy, if calculated using the color-spin interaction,
is negative although the spin-0 diquark is tighter~\cite{Jaffe:1999ze}.
Other possible diquarks with different structure, namely, $qq \in (J=1,\bar{\bm{3}}_c,\bm{6}_f), (J=0,\bm{6}_c,\bm{6}_f)$,
can be excluded from a possible building block because, first of all, their binding energy is repulsive and secondly
the resulting tetraquarks constructed from these diquarks predict the resonances with $I=3/2, 2$ which, however, have never been
observed in experiments.

The tetraquarks constructed from the spin-1 diquark, Eq.~(\ref{spin1diquark}), have the spin configuration,
\begin{equation}
|J,J_{12},J_{34}\rangle=|011\rangle\ ,
\end{equation}
and color structure, being $|\bm{1}_c,\bm{6}_c,\bar{\bm{6}}_c\rangle$, can be
written in a tensor notation as
\begin{eqnarray}
\frac{1}{\sqrt{96}} \Big( q^a q^b+q^b q^a \Big )
\Big (\bar{q}_a \bar{q}_b+\bar{q}_b \bar{q}_a\Big )\label{color wave function2}\ ,
\end{eqnarray}
in terms of individual quark colors.
Finally, since the diquark's flavor is still in $\bar{\bm{3}}_f$, the resulting tetraquarks form a {\it nonet} in flavor
whose wave function is again given by Eqs.~(\ref{flavor octet}),(\ref{flavor singlet}).

Then, we have two types of tetraquarks depending on the diquark being used in their construction.
Since the two types share the same
flavor, they can be labeled by the spin and color configurations as
$|000\rangle_{\bar{\bm{3}}_c,\bm{3}_c}$, $|011\rangle_{\bm{6}_c,\bar{\bm{6}}_c}$.
For notational simplicity,
we will suppress the color subscripts and denote the two types of tetraquarks simply by $|000\rangle$, $|011\rangle$, in the followings.

An important aspect that we want to address is that
the two types of tetraquark, $|000\rangle$, $|011\rangle$, can {\it mix} through the color-spin interaction.
Because of this mixing, the physical states can be identified by the eigenstates that diagonalize the hyperfine mass matrix
obtained from the expectation values of the color-spin interaction with respect to $|000\rangle$, $|011\rangle$.
If this mixing is strong, this can lead to a huge separation in hyperfine masses in the diagonal bases, which can explain
the huge mass gaps between the two nonets in PDG.

Indeed, this mixing framework tested in isovector members seems to work fine~\cite{Kim:2016dfq}.
By identifying the two eigenstates in isovector channel with $a_0 (980)$, $a_0 (1450)$,
the hyperfine mass splitting is found to reproduce the physical mass gap nicely.
This framework was also tested in the isodoublet members, $K_0^* (800)$, $K_0^* (1430)$,
with a more or less acceptable agreement.

In this work, we test this framework further in the isoscalar channel.
We have two resonances with $I=0$ in the light nonet, $f_0(500)$, $f_0(980)$,
and another two resonances in the heavy nonet, $f_0(1370)$, $f_0(1500)$.
Unlike the isovector and isodoublet cases~\cite{Kim:2016dfq}, these resonances
in the isoscalar channel can have the additional flavor mixing between the $I=0$ member in $\bm{8}_f$ and
the $I=0$ member belonging to $\bm{1}_f$.  We discuss this aspect in detail in Sec.~\ref{sec:spin0} below.

\section{Candidates for spin-1 and spin-2 tetraquarks}
\label{sec:spin12}

Our mixing framework for the spin-0 tetraquarks introduces the spin-1 diquark as an additional building block.
This means that the two nonets,
$a_0 (980)$, $K_0^* (800)$, $f_0 (500)$, $f_0 (980)$ in light mass,
and $a_0 (1450)$, $K_0^* (1430)$, $f_0 (1370)$, $f_0 (1500)$ in heavy mass,
can be generated by the mixing of the two spin configurations, $|000\rangle$ and $|011\rangle$.
If this scenario works, an immediate expectation is the existence of additional tetraquark nonets with spin-1 and spin-2
that can be constructed from the spin-1 diquark, Eq.~(\ref{spin1diquark}), also.
Their spin configurations should be
\begin{eqnarray}
|111\rangle\ ;\quad |211\rangle\ .
\end{eqnarray}
The color and flavor configurations are the same for the two nonets,
given by Eq.~(\ref{color wave function2}) for color, Eqs.~(\ref{flavor octet}),(\ref{flavor singlet}) for flavor.
We then ask what the corresponding resonances are in PDG.
The existence of the corresponding resonances may not be strictly enforced because of the possibility
that they can be hidden in the two-meson continuum.
Nevertheless, it may be interesting to search for possible candidates in PDG
with the quantum numbers, $J^P=1^+, 2^+$, which then can support our mixing framework more clearly.


\begin{table}
\centering
\begin{tabular}{c|c|c|c|c}  \hline\hline
$J^{PC}$ & $I$ &  Meson  & Mass (MeV) & $\Gamma$ (MeV) \\
\hline
\multicolumn{5}{c}{[Spin-1 resonances in PDG]} \\
\hline
\multirow{4}{*}{$1^{+-}$} & $ 0 $ & \underline{$h_1 (1170)$} & 1170 &  360   \\

         & $ 1 $ & \underline{$b_1 (1235)$} & 1229.5 &  142  \\
         & $ ? $ & \underline{$h_1 (1380)$} & 1386 &  91   \\
         & $ 0 $ & $h_1 (1595)$ & 1594 &  384   \\
\hline
\multirow{5}{*}{$1^{++}$} & $ 1 $ & $a_1 (1260)$ & 1230 &  250-600   \\
         & $ 0 $ & $f_1 (1285)$ & 1281.9 &  24.2   \\
         & $ 1 $ & $a_1 (1420)$ & 1414   & 153 \\
         & $ 0 $ & $f_1 (1420)$ & 1426.4 &  54.9  \\
         & $ 0 $ & $f_1 (1510)$ & 1518 &  73  \\
         & $ 1 $ & $a_1 (1640)$ & 1647 &  254 \\
\hline
\multirow{3}{*}{$1^{+}$}
        & $1/2$ & \underline{$K_1 (1270)$} & 1272 & 90  \\
        & $1/2$ & $K_1 (1400)$ & 1403 & 172  \\
        & $1/2$ & $K_1 (1650)$ & 1650 & 150  \\
\hline\hline
\multicolumn{5}{c}{[Spin-2 resonances in PDG]} \\
\hline
\multirow{11}{*}{$2^{++}$} & $ 0 $ & $f_2 (1270)$ & 1275.1 &  181.1   \\
         & $ 1 $ & $a_2 (1320)$ & 1318.3 &  105  \\
         & $ 0 $ & $f_2 (1430)$ & 1430 &  ?   \\
         & $ 0 $ & $f_2 (1525)$ & 1525 &  73   \\
         & $ 0 $ & $f_2 (1565)$ & 1562 &  134   \\
         & $ 0 $ & $f_2 (1640)$ & 1639 &  99   \\
         & $ 1 $ & $a_2 (1700)$ & 1732 &  194  \\
         & $ 0 $ & $f_2 (1810)$ & 1815 &  197  \\
         & $ 0 $ & $f_2 (1910)$ & 1903 &  196  \\
\cline{2-5}
         & \multicolumn{4}{l}{5 more with $I=0$, $f_2 (1950)$,$f_2 (2010)$,}\\
         & \multicolumn{4}{l}{$f_2 (2150)$, $f_2 (2300)$, $f_2 (2340)$}\\
\hline
\multirow{2}{*}{$2^{+}$}
        & $1/2$ & $K_2^* (1430)$  & 1425 & 98.5  \\
        & $1/2$ & $K_2^* (1980)$ & 1973 & 373  \\
\hline\hline
\end{tabular}
\caption{The upper part is the resonances with $J^{P}=1^+$ and the lower part is the resonances with $J^{P}=2^+$ collected from PDG.
The underlined members in the upper part
are possible candidates for the tetraquark nonet in spin-1. See the text for this selection.
}
\label{spin1,2 resonances}
\end{table}

To look for such candidates in spin-1, we have collected the resonances
with $J^P=1^+$ in the upper portion of Table~\ref{spin1,2 resonances}.
For the isodoublet channel, $K_1 (1270)$ can be a strong candidate because it fits to the mass ordering generated
from the hyperfine masses among spin-0 and spin-1 members~\cite{Kim:2016dfq}.
For isovector and isoscalar channels,
the additional quantum number, $C$-parity, which can be assigned to the $I_z=0$ members, can be used to narrow down
the possible candidates.  In Table~\ref{spin1,2 resonances}, for the $J=1$ case, the $I=0,1$ members are divided into
two categories depending on $C$-parity, one with $J^{PC}=1^{++}$ and the other with $J^{PC}=1^{+-}$.
On the other hand, one can directly determine the $C$-parity of the spin-1 tetraquarks
using the wave functions for them.
As demonstrated in Appendix, it can be shown that~\footnote{The similar proof can be found also in Erratum of Ref.~\cite{Kim:2016dfq}.
Thus, $a_1(1260)$, which was originally identified as the $I=1$ candidate for $|111\rangle$ in
Ref.~\cite{Kim:2016dfq}, needs to be replaced by $b_1(1235)$.  But since their experimental masses are almost the same,
$M[b_1(1235)] = 1229.5$ MeV, $M[a_1(1260)] = 1230$ MeV,
the discussion in that paper, which is mostly based on the mass splittings, is unaltered.} $C=-$.
Thus, the spin-1 candidates in isovector and isoscalar channels must be sought from the resonances with $J^{PC}=1^{+-}$.

For the isovector channel, we have only one resonance in PDG, $b_1(1235)$, and this can be a candidate for the spin-1 tetraquark.
For the isoscalar channel, we need two resonances to fill the spin-1 nonet and, based on their masses,
we choose $h_1 (1170)$ and $h_1 (1380)$
to be the candidates.  The other isoscalar member, $h_1(1595)$, seems to be too heavy to be a candidate.
Currently in PDG, the isospin of $h_1 (1380)$ is not determined yet even though its name assignment
seemingly indicates that this is an isoscalar resonance.  In this sense, this selection is not definite.

Another problem in this selection is that the experimental mass of $b_1(1235)$ is slightly smaller
than that of $K_1 (1270)$ by 40 MeV.  This violates the mass hierarchy discussed in Sec.~\ref{sec:motivation}, namely,
the isovector members are expected to be heavier than the isodoublet members.
But $b_1(1235)$, $K_1 (1270)$ have decay widths 140 MeV, 90 MeV respectively.  The decay widths are
rather large compared to
the mass gap so these resonances broaden by the decay width have some chance that the mass ordering is reversed.
Therefore, even though the selection for the candidates needs more clarification,
we can at least claim that, in PDG, there are some candidates in the $J^{P}=1^+$ channel to support our tetraquark mixing framework.

Possible candidates for the spin-2 tetraquarks can be selected from the resonances with $J^P=2^+$
shown in the bottom portion of Table~\ref{spin1,2 resonances}. The $C$-parity of the spin-2 tetraquark
can be proven to be even (see Appendix).
For the isodoublet member, $K_2^* (1430)$ can be chosen
to be the candidate even though its mass is rather small.  This selection forces us to choose the isovector member
to be $a_2 (1700)$ as it is heavier than $K_2^* (1430)$.  Unfortunately, these candidates fail to give the
mass splitting consistent with their hyperfine mass splitting~\cite{Kim:2016dfq}.  This failure might
be due to the small mass of the candidate $K_2^* (1430)$.  To achieve the consistency with the hyperfine spliting,
it is anticipated to have a spin-2 resonance with a mass around 1700 MeV in the isodoublet channel.

Also, for the isoscalar members in $2^{++}$, there are many resonances to choose from the lower part
of Table~\ref{spin1,2 resonances}. Their masses
are not well separated so the selection can be ambiguous.
Moreover, as we will see later, the hyperfine masses are positive for the spin-2 tetraquarks indicating that
the possible candidates are less bound than the spin-0,1 tetraquarks.
The possible candidates may even have a repulsive binding so they have more probability of being hidden in the two-meson continuum.

Under this circumstance, the selection tends to involve some arbitrariness.
Even if we come up with certain candidates that happen to yield some nice phenomenological consistency,
it may not be easy to justify the selection {\it a priori}.
With this reason, we do not look for the possible candidates for the spin-2 tetraquarks in this work.
However, this does not mean that there are no spin-2 tetraquarks to support the tetraquark mixing framework.
The problem is that we have too much ambiguity in selecting them.

\section{Hyperfine mass}
\label{sec:mass}

The tetraquark wave functions introduced in Sec.~\ref{sec:motivation},\ref{sec:spin12}
can be tested by comparing their theoretical masses calculated from the wave functions
with their experimental counterparts.
A hadron mass ($M_H$) can be estimated formally by
\begin{equation}
M_H=\sum_{i} m_i +\langle V \rangle\label{mass_f}\ ,
\end{equation}
where $m_i$ is the constituent mass of the $i$th quark and $\langle V \rangle$ is the expectation value
of the potential with respect to the hadron wave function.
The potential, $V$, which acts on constituent quarks, has two different sources,
one-gluon exchange potential~\cite{DeRujula:1975qlm,Keren07,Silve92,GR81} and the
instanton-induced interaction~\cite{OT89,Oka:1990vx}.
These two sources can be effectively parameterized as
\begin{eqnarray}
V =  \sum_{i < j}  v_0 J_i\cdot J_j \frac{\lambda_i \cdot \lambda_j}{m_i^{} m_j^{}}
+ \sum_{i < j} v_1^{} \frac{\lambda_i \cdot \lambda_j}{m_i^{} m_j^{}}  + v_2\ ,
\end{eqnarray}
where $\lambda_i$ denotes the Gell-Mann matrix for SU(3)$_c$, $J_i$ is the spin of the $i$th quark.
The first term represents the color-spin interaction, $V_{CS}$, and the second term the color-electric term, $V_{CE}$.
The parameters $v_0, v_1, v_2$ can be determined in principle by fitting hadron masses.
However, being an effective potential, its universal application can be limited in practice.

More reliable prediction can be made from the mass splitting.
The hyperfine masses, which are the expectation values of the color-spin interaction,
can be used for this purpose.
Specifically, as advocated in Refs.~\cite{Kim:2014ywa,Kim:2016tys,Kim:2016dfq},
the hyperfine mass splitting can approximate the mass difference of hadrons quite well,
\begin{equation}
\Delta M_H \approx \Delta \langle V_{CS} \rangle \ ,
\label{mass_splitting}
\end{equation}
as long as the difference
are taken for hadrons with the same flavor content and the same color configuration.
This mass relation works well especially for the lowest-lying baryons and
mesons~\footnote{See for example Table VI, VII in Ref.~\cite{Kim:2014ywa}.} because
the leading quark mass term in Eq.~(\ref{mass_f}),
which could be the biggest source of
uncertainty, cancels in the difference among hadrons with the same flavor content.
The color electric terms ($V_{CE} \sim \lambda_i \cdot \lambda_j$)
also cancel in the difference because they have the same color configuration.

In this work, we will use this mass formula, Eq.~(\ref{mass_splitting}),
to test our tetraquark wave functions.
In this application, one problem is whether the color electric term still cancels away in the mass difference
between the tetraquarks because our wave functions have two types of color
configurations, $|\bm{1}_c,\bar{\bm{3}}_c,\bm{3}_c\rangle$, $|\bm{1}_c,\bm{6}_c,\bar{\bm{6}}_c\rangle$.
In fact, in our actual calculation, we have included the color-electric term and
found that it gives negligible contributions
to the mass difference of Eq.~(\ref{mass_splitting}).  This aspect has been demonstrated explicitly
in the calculations of the isovector and
isodobulet resonances~\cite{Kim:2016dfq}. The similar thing applies to the isoscalar resonances here.
Therefore, in order to make our presentation more focusing, we do not discuss the color-electric term.

As for the input parameters in our calculation, we take the standard values for the constituent quark masses
$m_u = m_d = 330$~MeV, $m_s = 500$~MeV as
in our previous works~\cite{Kim:2014ywa,Kim:2016tys,Kim:2016dfq}.
For the strength $v_0$ of the color-spin interaction, we take the one determined from
the tetraquark framework developed for the $D$ meson excited states
where $v_0$ is fixed from the mass splitting of $D_0^*(2318)-D_2^*(2463)$~\cite{Kim:2014ywa}.

\section{Mixing in the spin-0 channel}
\label{sec:spin0}

As we discussed in Sec.~\ref{sec:motivation}, PDG has two nonets in spin-0 channel, light and heavy nonets, which, in our tetraquark model,
can be generated by the spin configuration mixing.
The isoscalar resonances of our concern are $f_0(500)$, $f_0(980)$ in the light nonet, and $f_0(1370)$, $f_0(1500)$
in the heavy nonet.
From the mass ordering, it is natural to consider that the spin configuration mixing relates $f_0(500)$, $f_0(1370)$ as they
are expected to be the same flavor member locating in the two different nonets.
Also the other two, $f_0(980)$ and $f_0(1500)$, which constitute a pair with higher masses,
are another members to be connected by the spin configuration mixing.
What makes the situation complicate in the isoscalar resonances is an additional flavor
mixing that generates $f_0(500)$, $f_0(980)$ in the light nonet, and $f_0(1370)$, $f_0(1500)$ in the heavy
nonet.  In this section, we introduce the flavor mixing first and then discuss how the spin configuration mixing can
be incorporated.

As shown in Fig.~\ref{flavor structure}, there are two members with $I=0$,
one member belonging to $\bm{1}_f$ and the other to $\bm{8}_f$.
If SU(3)$_f$ symmetry is exact, the two members
have the flavor structure as
\begin{eqnarray}
|\bm{1}_f\rangle_{I=0} &=& \frac{1}{\sqrt{3}} \Big \{ [ud][\bar{u}\bar{d}] + [ds][\bar{d}\bar{s}]+ [su][\bar{s}\bar{u}] \Big \} \label{singlet}\ ,\\
|\bm{8}_f\rangle_{I=0} &=& \frac{1}{\sqrt{6}} \Big \{ 2[ud][\bar{u}\bar{d}] - [ds][\bar{d}\bar{s}] - [su][\bar{s}\bar{u}] \Big \} \label{octet}\ .
\end{eqnarray}
Since the strange quark is heavier than $u$,$d$ quarks in the real world,
$|\bm{1}_f\rangle_{I=0}$ is heavier than $|\bm{8}_f\rangle_{I=0}$.
In the $J=0$ channel, we can match these two states,
($|\bm{8}_f\rangle_{I=0}$, $|\bm{1}_f\rangle_{I=0}$), to [$f_0(500)$, $f_0(980)$] in the light nonet,
and to [$f_0(1370)$, $f_0(1500)$] in the heavy nonet.
We call this case as ``SU(3) symmetric case'' (SSC) in this work.

However, these two states, ($|\bm{8}_f\rangle_{I=0}$, $|\bm{1}_f\rangle_{I=0}$), are expected
to mix in flavor according to Okubo-Zweig-Iizuka (OZI) rule.
This rule basically separates the parts containing strange quarks
in the wave functions from
the parts without strange quarks and it is originally applied successfully to the vector channel like $\omega$, $\phi$.
We believe that its generalization can be applied to multiquark systems,
like tetraquarks as well as pentaquarks~\cite{Lee:2004bsa}.

In the ideal mixing scenario of the generalized OZI rule, the separation becomes maximal and
the isoscalar resonances are represented by the flavor structure as
\begin{eqnarray}
|L\rangle &=&  [ud][\bar{u}\bar{d}]  \label{low}\ ,\\
|H\rangle &=& \frac{1}{\sqrt{2}} \{ [ds][\bar{d}\bar{s}] + [su][\bar{s}\bar{u}] \} \label{high}\ .
\end{eqnarray}
The notations, $|L\rangle$, $|H\rangle$, have been introduced in order to indicate that $|L\rangle$ is light and $|H\rangle$ is heavy in mass.
Again, if this scenario is realized in the real world, these ideal mixing states in the $J=0$ channel,
$\left ( |L\rangle, |H\rangle \right )$, can be matched to $\left [f_0(500), f_0(980) \right ]$ in the light nonet,
and to $\left [f_0(1370), f_0(1500) \right ]$ in the heavy nonet.
We call this situation as ``ideal mixing case''(IMC) in this work.

In general, the physical resonances may lie between the two extremes,
SSC and IMC, and we may write them as mixtures of the form,
\begin{eqnarray}
|\psi_1 \rangle &=& a |L\rangle + b |H\rangle  \label{psi1}\ ,\\
|\psi_2 \rangle &=& -b |L\rangle + a |H\rangle  \label{psi2}\ .
\end{eqnarray}
These constitute the general expressions from which one can recover the two limiting cases by
setting the flavor mixing parameters, $a,b$.
When $a=\sqrt{2/3}$, $b=-\sqrt{1/3}$, one gets the SSC,
\begin{eqnarray}
|\psi_1 \rangle \rightarrow |\bm{8}_f\rangle_{I=0} \ , |\psi_2 \rangle \rightarrow |\bm{1}_f\rangle_{I=0} \nonumber\ .
\end{eqnarray}
When $a=1$, $b=0$, one gets the IMC,
\begin{eqnarray}
|\psi_1 \rangle \rightarrow |L \rangle \ , |\psi_2 \rangle \rightarrow |H \rangle \nonumber\ .
\end{eqnarray}
In both limits, we see the mass ordering that $|\psi_1\rangle$ is lighter than $|\psi_2\rangle$.  This mass ordering is
maintained as long as $a, b$ vary within the two limiting cases~\footnote{Once $a$ is chosen, $b$ is determined from the normalization
condition, $a^2+b^2=1$. Our sign convention for $b$ is $b=-\sqrt{1-a^2}$ and this is consistent with the sign of the parameters in the SSC.}.
Again, the mass ordering in the $J=0$ channel leads us to identify the two states, ($|\psi_1\rangle$, $|\psi_2\rangle$),
as [$f_0(500)$, $f_0(980)$] in the light nonet,
and to [$f_0(1370)$, $f_0(1500)$] in the heavy nonet.
Later, we will fix the flavor mixing parameters, $(a,b)$, by equating hyperfine mass splitting to the physical mass
splitting between $f_0 (980)$, $f_0(1500)$
after including spin configuration mixing.
This result based on this fitting will be referred to ``realistic case with fit'' (RCF) in this work.

Each state introduced in this section,  $|\psi_1\rangle$, $|\psi_2\rangle$ in RCF,
$|\bm{8}_f\rangle_{I=0}$, $|\bm{1}_f\rangle_{I=0}$ in SSC, $|L\rangle$, $|H\rangle$ in IMC,
can have all the spin configurations advocated in Sec.~\ref{sec:motivation}, \ref{sec:spin12}.
For example, $|\psi_1\rangle$ can be either $|\psi_1, 000\rangle$ or $|\psi_1, 011\rangle$ in $J=0$,
$|\psi_1,111\rangle$ in $J=1$, $|\psi_1,211\rangle$ in $J=2$.
Since we have two spin configurations in $J=0$, the spin-0 tetraquarks are also subject
to the spin configuration mixing in addition to the flavor mixing of the type Eqs.~(\ref{psi1}), (\ref{psi2}).

We now explain how the spin configuration mixing in the $J=0$ channel can be implemented
in $|\psi_2, 000\rangle$, $|\psi_2, 011\rangle$. The same prescription can be applied to $|\psi_1, 000\rangle$, $|\psi_1, 011\rangle$ similarly.
First, we write down Eq.~(\ref{psi2}) for the two spin configurations as
\begin{eqnarray}
|\psi_2, 000\rangle = -b |L, 000\rangle + a |H, 000\rangle \label{psi2spin000}\ ,\\
|\psi_2, 011\rangle = -b |L, 011\rangle + a |H, 011\rangle \label{psi2spin011}\ ,
\end{eqnarray}
where we have indicated the spin configurations for the ideal mixing states similarly.

As advocated in Sec.~\ref{sec:motivation}, these two spin configurations are expected to mix strongly
through the color-spin interaction.
The matrix elements of the color-spin interaction with respect to
the bases $|\psi_2, 000\rangle$, $|\psi_2, 011\rangle$, namely the hyperfine masses,
can be written in terms of those with respect to ideal mixing states as
\begin{widetext}
\begin{eqnarray}
\langle \psi_2,000|V_{CS}|\psi_2,000 \rangle &=& b^2 \langle L,000|V_{CS}|L,000 \rangle + a^2 \langle H,000|V_{CS}|H,000 \rangle \label{v11}\ ,\\
\langle \psi_2,011|V_{CS}|\psi_2,011 \rangle &=& b^2 \langle L,011|V_{CS}|L,011 \rangle + a^2 \langle H,011|V_{CS}|H,011 \rangle \label{v22}\ ,\\
\langle \psi_2,000|V_{CS}|\psi_2,011 \rangle &=& b^2 \langle L,000|V_{CS}|L,011 \rangle + a^2 \langle H,000|V_{CS}|H,011 \rangle \label{v12}\ .
\end{eqnarray}
\end{widetext}
Note that the ideal mixing states do not mix through
$V_{CS}$, namely $\langle L|V_{CS}|H \rangle=0$ because $|L \rangle$ $|H \rangle$ are orthogonal
in flavor space and $V_{CS}$ is blind on flavor.
Table~\ref{V for spin0} provides the numerical values for the matrix elements
involving the ideal mixing states $|L,000\rangle$, $|L,011\rangle$, $|H,000\rangle$, $|H,011\rangle$.
These are calculated by the general formulas
given in Table 2 of Ref.~\cite{Kim:2016dfq} and summing over the flavor combinations according to
Eqs.~(\ref{low}),(\ref{high}).  Before we move on, it is important to point out that the off-diagonal components between
the two spin configurations, $|000\rangle$, $|011\rangle$, in Table~\ref{V for spin0},
are comparable in magnitude with the diagonal elements.  This indicates that the spin configuration mixing is very strong.
Anyway, once the parameters, $a,b$, are chosen, one can determine
the numerical values for the hyperfine masses with respect to the states $|\psi_2,000 \rangle$, $|\psi_2,011 \rangle$.


\begin{table}[t]
\centering
\begin{tabular}{c|c}  \hline\hline
$|L\rangle$ with $J=0$                         & $|H\rangle$ with $J=0$ \\
\hline
$\langle L,000|V_{CS}|L,000 \rangle = -263.46$ & $\langle H,000|V_{CS}|H,000 \rangle = -173.88$ \\
$\langle L,000|V_{CS}|L,011 \rangle = -322.67$ & $\langle H,000|V_{CS}|H,011 \rangle = -222.29$ \\
$\langle L,011|V_{CS}|L,011 \rangle = -483.01$ & $\langle H,011|V_{CS}|H,011 \rangle = -331.48$ \\
\hline\hline
\end{tabular}
\caption{Numerical values for $\langle V_{CS}\rangle$ are presented here
for the ideal mixing states with the specified spin configurations in the $J=0$ channel.
All the numbers are given in MeV unit.}
\label{V for spin0}
\end{table}

Because of the mixing elements, the hyperfine masses form a $2\times2$ matrix in the bases
$|\psi_2,000 \rangle$, $|\psi_2,011 \rangle$.
The physical states are the eigenstates that diagonalize this hyperfine matrix, and they are of course
mixtures of $|\psi_2,000 \rangle$, $|\psi_2,011 \rangle$.
They can be identified
as the isoscalar resonances, $f_0(980)$ and $f_0(1500)$.
This means that
\begin{eqnarray}
&&|f_0(1500)\rangle = -\alpha_2 |\psi_2, 000 \rangle + \beta_2 | \psi_2, 011 \rangle \nonumber\ ,\\
&&|f_0(980) \rangle = \beta_2 | \psi_2, 000 \rangle + \alpha_2 | \psi_2, 011 \rangle \label{mixing2} \ ,
\end{eqnarray}
with the mixing parameters, $\alpha_2$, $\beta_2$, to be determined by the diagonalization.
The similar mixing with the spin-1 diquark configuration
was also reported in Ref.~\cite{Jaffe04, Black:1998wt} where this mixing
was used to explain the small masses of the lowest-lying states in the $0^+$ channel
without identifying the other states with higher masses.
These parameters,
$\alpha_2$, $\beta_2$, are functions of the flavor mixing parameters $a,b$.
The eigenvalues, which are the hyperfine masses in the physical bases,
$\langle f_0(980)| V_{CS} |f_0(980)\rangle$, $\langle f_0(1500)| V_{CS} | f_0(1500) \rangle$, are also
the functions of the parameters $a,b$.  Therefore, once $a,b$ are given, we can determine all the terms needed
in our analysis.

What is interesting in Eq.~(\ref{mixing2}) is that the relative signs between $|\psi_2, 000 \rangle$, $| \psi_2, 011 \rangle$ are opposite
in the two equations.  This sign difference can make a clear distinction in the fall-apart decays of
$|f_0(1500)\rangle$, $|f_0(980)\rangle$.
Namely, the couplings associated with their two-meson decays are enhanced in one resonance
while they are suppressed in the other resonance.
Later, we will discuss the consequences of this interesting aspect further in Sec.~\ref{sec:fall apart}.

The flavor mixing parameters fixed in the two limiting cases are
$a=\sqrt{2/3}$, $b=-\sqrt{1/3}$ in the SSC, and $a=1$, $b=0$ in the IMC. The corresponding hyperfine masses are
calculated to be
\begin{eqnarray}
\begin{array}{l|c|c}
     &   \text{SSC} & \text{IMC}  \\
\hline 
  \langle f_0(1500)| V_{CS} | f_0(1500) \rangle   & -22.03      &-16.84     \\
   \langle f_0(980)| V_{CS} |f_0(980)\rangle   & -563.7    & -488.52
\end{array}
\nonumber\ ,
\end{eqnarray}
in MeV unit.
We can clearly see that the separation in hyperfine masses is huge, around 500 MeV, and we emphasize that this
is mainly driven by the strong mixing between the two spin configurations, $|\psi_2, 000\rangle$, $|\psi_2, 011\rangle$.

According to our mass formula, Eq.~(\ref{mass_splitting}), the hyperfine mass splitting,
$\Delta \langle V_{CS} \rangle=\langle f_0(1500)| V_{CS} | f_0(1500) \rangle-\langle f_0(980)| V_{CS} |f_0(980)\rangle$, needs
to be equated to the mass splitting $\Delta M_{H}= M[f_0(1500)]-M[f_0(980)]$ if our tetraquark model works.
The calculated values of the hyperfine mass splitting, $\Delta \langle V_{CS} \rangle$, are
\begin{eqnarray}
\Delta \langle V_{CS} \rangle= 541.7~\text{MeV}\text{(SSC)},~471.7~\text{MeV}\text{(IMC)}\label{hmass splitting2}\ .
\end{eqnarray}
These numbers are relatively close to the experimental mass splitting, $\Delta M_{H}= 515$ MeV.
Since the realistic case is expected to lie between the two limits, SSC and IMC, we may claim that
the tetraquark mixing model works for the resonances $f_0(980)$, $f_0(1500)$ very well.

For the realistic situation, we need to determine the parameters $a,b$ using the experimental inputs.
To do this, we rely on the mass formula, Eq.~(\ref{mass_splitting}), using the experimental masses of $f_0(980)$, $f_0(1500)$ as inputs.
By tuning the parameter $a$ from $a=\sqrt{2/3}$ (SSC) to $a=1$ (IMC),
we numerically look for its value that leads to the hyperfine mass splitting equivalent to the
experimental mass splitting, 515 MeV.  Using that value of $a$, the other parameter $b$ is fixed to be $b=-\sqrt{1-a^2}$.
This fitting process leads to the RCF parameters,
\begin{eqnarray}
a=0.8908, ~~b=-0.4543\label{RCF par}\ .
\end{eqnarray}

The same prescription can be applied to $|\psi_1\rangle$.
First, according to Eq.~(\ref{psi1}),  the spin configurations for $|\psi_1\rangle$ are related to the ideal mixing
states as
\begin{eqnarray}
|\psi_1, 000\rangle = a |L, 000\rangle + b |H, 000\rangle \label{psi1spin000}\ ,\\
|\psi_1, 011\rangle = a |L, 011\rangle + b |H, 011\rangle \label{psi1spin011}\ .
\end{eqnarray}
Since $|\psi_1\rangle$ is obtained from $|\psi_2\rangle$ simply by replacing $-b\rightarrow a, a\rightarrow b$,
the corresponding hyperfine formulas for the states $|\psi_1,000 \rangle$, $|\psi_1,011 \rangle$
can be obtained from Eqs.~(\ref{v11}),(\ref{v22}),(\ref{v12}) by replacing $\psi_2\rightarrow \psi_1, b^2 \leftrightarrow a^2$.
Again, the eigenstates are mixtures of $|\psi_1,000 \rangle$, $|\psi_1,011 \rangle$
and they should represent the isoscalar resonances with lighter masses, $f_0(500)$ and $f_0(1370)$.
This means that
\begin{eqnarray}
&&|f_0(1370)\rangle = -\alpha_1 |\psi_1, 000 \rangle + \beta_1 | \psi_1, 011 \rangle \nonumber\ ,\\
&&|f_0(500) \rangle = \beta_1 | \psi_1, 000 \rangle + \alpha_1 | \psi_1, 011 \rangle \label{mixing1} \ .
\end{eqnarray}
Depending on the three cases, SSC, IMC, RCF,
our results for the hyperfine mass for $f_0(1370)$, $f_0(500)$ in MeV unit are
\begin{eqnarray}
\begin{array}{l|c|c|c}
     &   \text{SSC} & \text{IMC} & \text{RCF} \\
\hline 
\langle f_0(1370)| V_{CS} | f_0(1370) \rangle  & -27.22      &-32.4     & -29.19 \\
\langle f_0(500)| V_{CS} | f_0(500) \rangle & -638.88     & -714.07  & -667.51
\end{array}
\nonumber\ .
\end{eqnarray}

For $f_0 (1370)$ and $f_0 (500)$, we find the hyperfine mass splitting,
$\Delta \langle V_{CS} \rangle=\langle f_0(1370)| V_{CS} | f_0(1370) \rangle-\langle f_0(500)| V_{CS} |f_0(500)\rangle$ as
\begin{eqnarray}
  \Delta \langle V_{CS} \rangle &=& 611.66~\text{MeV}\text{(SSC)},~681.67~\text{MeV}\text{(IMC)}, \nonumber\\
  && 638.32~\text{MeV}\text{(RCF)}
\label{hmass splitting1}\ ,
\end{eqnarray}
depending on the three cases.
Once again, we have huge mass gap generated from the spin configuration mixing.
This gap is also insensitive to the three different cases being considered, 10 \% or less.
But currently it is not clear whether this result agrees with the experimental mass splitting because
the masses of $f_0(1370)$, $f_0(500)$ are not fixed well experimentally.
As one can see in Table~\ref{spin0++}, $M[f_0(1370)]$ is given in the range, 1200-1500 MeV, and $M[f_0(500)]$ is
in the range, 400-550 MeV~\footnote{This is in fact
the reason why $f_0(1370)$, $f_0(500)$ are not used to determine the parameters $a,b$ in this work.}. Furthermore,
their decay widths are very large, $\Gamma[f_0(1370)]=200 - 500$ MeV, $\Gamma[f_0(500)]=400 - 700$ MeV.
Nevertheless, if we take their central values of the given mass ranges, one can crudely estimate the experimental mass
splitting, $\Delta M_{H}= M[f_0(1370)]-M[f_0(500)]=875$ MeV
which is about 200 MeV larger than the hyperfine mass splitting.
Of course, our hyperfine mass splitting does not necessarily agree with this crude mass splitting
but, from this comparison, we can see at least a tendency that the huge separation in hyperfine masses
qualitatively matches with the
actual mass splitting.

Before closing this section, we present in Table~\ref{parameters} the numerical values for the mixing parameters,
$\alpha_{1}, \alpha_{2}$ and $\beta_{1}, \beta_{2}$,
appearing in Eqs.~(\ref{mixing2}), (\ref{mixing1}) depending on the flavor mixing parameters, $a,b$,
in the three different cases, SSS, IMC, RCF. As one can see in the table,
the configuration mixing parameters
are almost insensitive to the three different cases.  In fact, their values approximately satisfy,
$\alpha_1\approx \alpha_2 \approx \sqrt{2/3}$, $\beta_1\approx \beta_2 \approx \sqrt{1/3}$.
It is very interesting to notice that, since $\alpha_1 > \beta_1$, $\alpha_2 > \beta_2$ in Eqs.~(\ref{mixing2}), (\ref{mixing1}),
the members in the light nonet, $f_0(500)$, $f_0(980)$, have more probability to stay in the spin configuration
$|\psi_1, 011\rangle$ than in $|\psi_1, 000\rangle$.  This is very different from the common expectation
that the spin-0 diquark configuration is dominant in the formation of tetraquarks.

\begin{table}[t]
\centering
\begin{tabular}{l|c|c|c|c|c|c}\hline\hline
        &  $a$     & $b$      & $\alpha_1$ & $\beta_1$ & $\alpha_2$ & $\beta_2$\\
\hline
SSC     & 0.8165   &-0.5774   & 0.8140     & 0.5809   & 0.8152     & 0.5792     \\
IMC     & 1        & 0        & 0.8130     & 0.5822   & 0.8167     & 0.5770    \\
RCF     & 0.8908   & -0.4543  & 0.8136     & 0.5814   & 0.8157     & 0.5784    \\
\hline\hline
\end{tabular}
\caption{Here are the mixing parameters depending on the three different cases, SSC, IMC, RFC.
The parameters associated with the spin configuration
mixing are found to be almost insensitive to the cases.}
\label{parameters}
\end{table}

\section{Hyperfine masses for the spin-1, 2 tetraquarks}
\label{sec:hf_spin12}

We now discuss the hyperfine masses for isoscalar tetraquarks in the $J=1,2$ channels.
Unlike to the $J=0$ case, there is no spin configuration mixing. So we need to consider
the flavor mixing only.  This means that, in these spin channels, the physical states
can be directly matched to the $J=1,2$ counterparts of Eqs.~(\ref{psi1}),(\ref{psi2}).
In other words, the physical states in $J=1$ are given by
\begin{eqnarray}
|\psi_1, 111 \rangle &=& a |L, 111\rangle + b |H, 111\rangle  \label{psi11}\ ,\\
|\psi_2, 111 \rangle &=& -b |L, 111\rangle + a |H, 111\rangle  \label{psi21}\ ,
\end{eqnarray}
with the mass ordering that $|\psi_2, 111 \rangle$ is heavier than $|\psi_1, 111 \rangle$.
The physical states in $J=2$ are
\begin{eqnarray}
|\psi_1, 211 \rangle &=& a |L, 211\rangle + b |H, 211\rangle  \label{psi12}\ ,\\
|\psi_2, 211 \rangle &=& -b |L, 211\rangle + a |H, 211\rangle  \label{psi22}\ .
\end{eqnarray}
Since the flavor structures are the same as in the $J=0$ case, we may use
the same parameters, $a,b$, determined in the three different cases above. Namely, we take
$a=\sqrt{2/3}$, $b=-\sqrt{1/3}$ for the SSC, $a=1$, $b=0$ for the IMC and
$a=0.8908$, $b=-0.4543$ for the RCF [see Eq.~(\ref{RCF par})].

To calculate the mass splitting using the mass formula, Eq.~(\ref{mass_splitting}),
we again need to evaluate the hyperfine masses with respect to these states.
These are related to the hyperfine masses in the ideal mixing states through
\begin{widetext}
\begin{eqnarray}
\langle \psi_1,111|V_{CS}|\psi_1,111 \rangle &=& a^2 \langle L,111|V_{CS}|L,111 \rangle + b^2 \langle H,111|V_{CS}|H,111 \rangle \label{hmass11}\ ,\\
\langle \psi_2,111|V_{CS}|\psi_2,111 \rangle &=& b^2 \langle L,111|V_{CS}|L,111 \rangle + a^2 \langle H,111|V_{CS}|H,111 \rangle \label{hmass21}\ ,\\
\langle \psi_1,211|V_{CS}|\psi_1,211 \rangle &=& a^2 \langle L,211|V_{CS}|L,211 \rangle + b^2 \langle H,211|V_{CS}|H,211 \rangle \label{hmass12}\ ,\\
\langle \psi_2,211|V_{CS}|\psi_2,211 \rangle &=& b^2 \langle L,211|V_{CS}|L,211 \rangle + a^2 \langle H,211|V_{CS}|H,211 \rangle \label{hmass22}\  .
\end{eqnarray}
\end{widetext}
The hyperfine masses in the ideal mixing bases, $|L\rangle$, $|H\rangle$, can be
calculated similarly as before, i.e., by summing over flavor combinations given in Eqs.~(\ref{low}),(\ref{high})
using the general formulas provided in Table 2 of Ref.~\cite{Kim:2016dfq}.
Their values are listed in the top portion of Table~\ref{V for spin 1 and 2}.
Plugging them in Eqs.~(\ref{hmass11}), (\ref{hmass21}), (\ref{hmass12}), (\ref{hmass22}), we obtain the hyperfine masses
in the physical bases calculated for the three cases, SSC, IMC, RCF. Their numerical values are listed in the bottom portion of
Table~\ref{V for spin 1 and 2}.

Interestingly, the hyperfine masses in physical bases are positive for the $J=2$ channel while they are
negative for the $J=1$ channel. Thus, the spin-2 tetraquarks are expected to be either unbound or less bound than
the spin-0 and spin-1 tetraquarks.  Currently there are some arbitrariness in selecting the possible candidates for
the spin-2 tetraquarks from PDG.  In addition, due to the positive hyperfine masses, there is some possibility
that they may be hidden in the two-meson continuum.  Moreover, some test on the
isodoublet channel in Ref.~\cite{Kim:2016dfq} seems not conclusive.
Therefore, as advertised before,  we do not look for candidates
for the spin-2 tetraquarks in the present work.


\begin{table}[t]
\centering
\begin{tabular}{l|c|c|c}  \hline\hline
  Hyperfine masses   &   SSC & IMC & RCF \\
\hline 
$\langle L,111|V_{CS}|L,111 \rangle$ & - & -263.46 & - \\
$\langle H,111|V_{CS}|H,111 \rangle$ & - & -180.23 & - \\
$\langle L,211|V_{CS}|L,211 \rangle$ & - &  175.64 & - \\
$\langle H,211|V_{CS}|H,211 \rangle$ & - &  122.27 & - \\
\hline
$\langle \psi_1,111|V_{CS}|\psi_1,111 \rangle$  & -235.72      &-263.46    & -246.28 \\
$\langle \psi_2,111|V_{CS}|\psi_2,111 \rangle$  & -207.97      &-180.23    & -197.41 \\
$\langle \psi_1,211|V_{CS}|\psi_1,211 \rangle$  & 157.85       & 175.64    & 164.62 \\
$\langle \psi_2,211|V_{CS}|\psi_2,211 \rangle$  & 140.06       & 122.27    & 133.28 \\
\hline\hline
\end{tabular}
\caption{The top portion shows the numerical values of $\langle V_{CS}\rangle$ with respect
to the ideal mixing states in the $J=1,2$ channels.
The bottom portion provides the hyperfine masses
in the physical bases calculated from the top portion through Eqs.~(\ref{hmass11}), (\ref{hmass21}), (\ref{hmass12}), (\ref{hmass22})
using the mixing parameters, $a,b$, corresponding to the three different cases of flavor mixing, SSC, IMC, RCF.
All the numbers are given in MeV unit.}
\label{V for spin 1 and 2}
\end{table}

For the spin-1 channel, the isoscalar candidates are $h_1 (1170)$,
$h_1 (1380)$ as discussed in Sec.~\ref{sec:spin12}.  These resonances can be matched to our spin states, Eqs.~(\ref{psi11}),(\ref{psi21}), as
\begin{eqnarray}
|h_1 (1170)\rangle &=& |\psi_1, 111 \rangle  \label{h1_psi11}\ ,\\
|h_1 (1380)\rangle &=& |\psi_2, 111 \rangle  \label{h1_psi21}\ .
\end{eqnarray}
In Table~\ref{splitting in spin12}, we present the hyperfine mass splittings involving the spin-1 tetraquarks.
In order to test their reliability through the mass formula, Eq.~(\ref{mass_splitting}),
we also show the experimental mass splitting of $h_1 (1170)$ from the corresponding members in the
spin-0 nonets, $f_0 (500), f_0 (1370)$,
and the mass splitting of $h_1 (1380)$ from $f_0 (980), f_0 (1500)$. Note, the mass splitting between $h_1 (1170)$
and $h_1 (1380)$ can not be estimated from Eq.~(\ref{mass_splitting}) as they have different flavor content.
For completeness, the results for the spin-0 tetraquarks given in Sec.~\ref{sec:spin0} are also listed in the table.
We again notice that the hyperfine mass splitting is somewhat insensitive to the three different cases being considered.
But in comparison with the experimental mass splitting, we have only rough agreement.
Specifically, both splittings agree relatively well for $ h_1 (1170) - f_0 (1370)$, $ h_1 (1380) - f_0 (980)$
but their agreement is not so great for $ h_1 (1170) - f_0 (500)$, $ h_1 (1380) - f_0 (1500)$.
However, one has to remember that the experimental mass splitting can not be precise due to barely known masses
of $f_0 (500)$, $f_0 (1370)$ as well as their broad widths.
Thus, the precise agreement is not strictly anticipated in the present situation.
From this table, we can at least claim that there is a qualitative trend of matching where
the hyperfine mass splitting goes along with the experimental
mass splitting.  Therefore, even though the statement can not be made conclusive, we
have some signatures to support the spin-1 tetraquark picture.


\begin{table}
\centering
\begin{tabular}{l|c|c|c|c}  \hline\hline
Participating               & $\Delta M_{\text{exp}}$ & \multicolumn{3}{c}{$\Delta \langle V_{CS} \rangle$ (MeV)} \\
\cline{3-5}
resonances                  &  (MeV)                  &  SSC & IMC &  RCF \\
 \hline
\multicolumn{4}{c}{[For $f_0(500), f_0(1370), h_1 (1170)$]} \\
\hline
$ f_0(1370) - f_0(500)$     &  875                    & 611.7  & 681.7 & 638.3  \\[0.3mm]
$ h_1 (1170) - f_0 (500)$   & 695                     & 403.2  & 450.6 & 421.2 \\[0.3mm]
$ h_1 (1170) - f_0 (1370)$  & -180                    & -208.5 & -231.1 & -217.1\\[0.3mm]
\hline
\multicolumn{4}{c}{[For $f_0 (980), f_0 (1500), h_1 (1380)$]} \\
\hline
$ f_0 (1500) - f_0 (980)$   &  515                     & 541.7  & 471.7 &  \underline{515} \\[0.3mm]
$ h_1 (1380) - f_0 (980)$   & 396                      & 355.7  & 308.3 & 337.7   \\[0.3mm]
$ h_1 (1380) - f_0 (1500)$  & -119                     & -185.9 & -163.4 & -177.4 \\[0.3mm]
\hline\hline
\end{tabular}
\caption{The hyperfine mass splittings in the spin-0,1 channels are compared with the corresponding mass splittings
with the identification of the isocalar resonances, $f_0 (500), f_0 (1370), h_1 (1170)$
in the light members from the three different nonets,
and $f_0 (980), f_0 (1500), h_1 (1380)$ in the heavy members.  In calculating the experimental mass splitting
from $f_0(500)$, $f_0 (1370)$, we take the central values of their mass ranges given in PDG, namely,
$M[f_0(500)]=475$ MeV, $M[f_0(1370)]=1350$ MeV.
The three different results on $\Delta \langle V_{CS} \rangle$ are obtained depending on the flavor mixing parameters, $a,b$.
$\Delta \langle V_{CS} \rangle$ for $f_0(1500)- f_0(980)$ in RCF
is underlined in order to indicate that this value is fitted to reproduce $\Delta M_{\text{exp}}$, 515 MeV, by tuning the parameter, $a$.
}
\label{splitting in spin12}
\end{table}

\section{Fall-apart modes of the tetraquarks}
\label{sec:fall apart}

\begin{figure}[t]
\centering
\epsfig{file=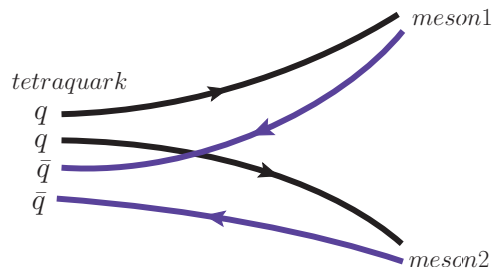, width=0.75\hsize}
\caption{A schematic diagram representing fall-apart decay. }
\label{fall apart}
\end{figure}

Tetraquarks have a unique decay mechanism called fall-apart decay~\cite{Jaffe77b}.
In this mechanism, quarks and antiquarks
inside a tetraquark are recombined into the two quark-antiquark pairs which then simply fall apart into two mesons if
the phase space is available.
A schematic view of this decay is shown in Fig.~\ref{fall apart}.
This mechanism is very different from a quark-antiquark system where
its decay proceeds through a creation of a quark-antiquark pair from the vacuum.
The fall-apart mechanism is expected to dominate in the multiquark systems
and it can be used to study the decay patterns of tetraquarks as well as pentaquarks~\cite{Lee:2004bsa}.
In this section, we study the fall-apart modes of the tetraquarks and investigate whether those
are consistent with the experimental decay modes of the corresponding resonances.
A similar formulation can be found for the isovector resonances,
$a_0(980)$, $a_0(1450)$, in Ref.~\cite{Kim:2017yur}.

\subsection{For spin-0 tetraquarks}

In the spin-0 channel, we have the isoscalar pair with light mass, $f_0(500)$, $f_0(1370)$, and
another pair with heavy mass, $f_0(980)$, $f_0(1500)$.
They can be expressed by linear combinations of
the ideal mixing states, $|L\rangle$, $|H\rangle$, according to Eqs.~(\ref{psi1spin000}),(\ref{psi1spin011}),(\ref{mixing1})
for the former, and Eqs.~(\ref{psi2spin000}),(\ref{psi2spin011}),(\ref{mixing2}) for the latter.
To examine the fall-apart modes of $f_0(500)$, $f_0(1370)$, $f_0(980)$ and $f_0(1500)$,
first we need to study the fall-apart modes of $|L\rangle$, $|H\rangle$.
Due to the kinematical accessibility, we are interested in their decay channels particularly into
two pseudoscalar mesons in this work.

The ideal mixing state, $|L\rangle$, having the flavor structure of $[ud][\bar{u}\bar{d}]$, Eq.(\ref{low}),
can be rearranged in terms of the quark-antiquark bases by combining
the first quark and third antiquark into one pair, which we call the (13) pair, and the second quark and
fourth antiquark into another pair, which we call the (24) pair. Under this regrouping, $|L\rangle$ can be written as
\begin{eqnarray}
[ud][\bar{u}\bar{d}] \doteq (u\bar{u})(d\bar{d})-(u\bar{d})(d\bar{u})\ .
\label{regrouping_L}
\end{eqnarray}
Since the lowest-lying pseudoscalar resonances in $J^P=0^+$ form a nonet in SU(3)$_f$,
the quark-antiquark pairs above have strong overlaps with the corresponding mesons as
\begin{eqnarray}
u\bar{u} &\Rightarrow& \frac{1}{\sqrt{3}} \eta_1 + \frac{1}{\sqrt{6}} \eta_8+ \frac{1}{\sqrt{2}} \pi^0\ ,\\
d\bar{d} &\Rightarrow& \frac{1}{\sqrt{3}} \eta_1 + \frac{1}{\sqrt{6}} \eta_8 - \frac{1}{\sqrt{2}} \pi^0\ ,\\
u\bar{d} &\Rightarrow& \pi^+\ ; \quad  d\bar{u} \Rightarrow \pi^-\ .
\end{eqnarray}
We ignore the $\eta-\eta^\prime$ mixing in this qualitative analysis.
Applying these replacements in Eq.~(\ref{regrouping_L}), one can readily obtain the fall-apart modes of $|L\rangle$
\begin{eqnarray}
|L\rangle \Rightarrow \frac{1}{3} \eta_1 \eta_1+ \frac{\sqrt{2}}{3} \eta_1\eta_8+ \frac{1}{6} \eta_8\eta_8
           - \frac{1}{2}\bm{\pi}\cdot\bm{\pi}\ .
\label{modes_L}
\end{eqnarray}
Because the matchings are not fully saturated by the specified mesons, this replacement needs to be understood up to an overall constant.

The same prescription can be applied to the other ideal mixing state, $|H\rangle$, with
the flavor structure $\frac{1}{\sqrt{2}} \{ [ds][\bar{d}\bar{s}] + [su][\bar{s}\bar{u}] \}$, Eq.(\ref{high}).
Here we simply quote the final expression for the fall-apart modes of $|H\rangle$,
\begin{eqnarray}
|H\rangle \Rightarrow \frac{1}{\sqrt{2}}\left\{\frac{2}{3} \eta_1 \eta_1- \frac{\sqrt{2}}{3} \eta_1\eta_8 - \frac{2}{3} \eta_8\eta_8
          - \overline{K} K \right \} ,
\label{modes_H}
\end{eqnarray}
where the pseudoscalar isodoublets are defined by
\begin{align}
\overline{K}=(K^-,\overline{K}^0)\ ,~   K &= \begin{pmatrix}
                      K^+ \\
                      K^0 \\
         \end{pmatrix}\label{isodoublets}\ .
\end{align}

Of course, there should be additional factors coming from the spin and color parts when we rearrange
the tetraquarks in terms of the (13)-(24) pairs, which then fall apart into two pseudoscalar mesons.
First, both (13)-(24) pairs need to be in a spin-0 state separately. From the spin configurations
of $|000\rangle$ and $|011\rangle$,
it is straightforward to extract the component with spins $J_{13}=J_{24}=0$,
\begin{eqnarray}
| 0 0 0 \rangle \rightarrow \frac{1}{2} | 0 0\rangle_{13} | 0 0\rangle_{24}\ ;\  | 0 1 1 \rangle \rightarrow \frac{\sqrt{3}}{2}
 | 0 0\rangle_{13} | 0 0\rangle_{24}.
\label{spin12}
\end{eqnarray}
Here, the notation, for example, $|0 0\rangle_{13}$ in the right-hand side,
denotes that $J_{13}=0$ and its spin projection, $(J_{13})_z=0$.

The color structures of $|000\rangle$, $|011\rangle$ are $|\bm{1}_c,\bar{\bm{3}}_c,\bm{3}_c\rangle$,
$|\bm{1}_c,\bm{6}_c,\bar{\bm{6}}_c\rangle$ respectively.
Now, for the color factors, we need to calculate the component when
both (13)-(24) pairs are combined into a color singlet separately.
From their tensor expressions in Eqs.~(\ref{color wave function1}),(\ref{color wave function2}), we find
\begin{eqnarray}
&&\frac{1}{\sqrt{12}}  \varepsilon_{abd}^{} \ \varepsilon^{aef}
\Big [ q_1^b q_2^d \Big ]
\Big [ \bar{q}^3_e \bar{q}^4_f \Big ]\rightarrow  \frac{1}{\sqrt{3}} \bm{1}_{c13} \bm{1}_{c24}\ ,\\
&&\frac{1}{\sqrt{96}} \Big [q_1^a q_2^b+q_1^b q_2^a \Big ]
\Big [\bar{q}^3_a \bar{q}^4_b+\bar{q}^3_b \bar{q}^4_a\Big ]
\rightarrow  \sqrt{\frac{2}{3}} \bm{1}_{c13} \bm{1}_{c24}\ .
\end{eqnarray}
Here we have enumerated the quark fields by numeric indices to show the grouping more clearly.
So, for instance, $\bm{1}_{c13}$ denotes the state where the (13) pair is in the color singlet.

The spin and color factors affect the fall-apart modes of Eqs.~(\ref{modes_L}),(\ref{modes_H})
that can be written symbolically by
\begin{eqnarray}
|L,000\rangle &\Rightarrow& \frac{1}{2\sqrt{3}}|L\rangle\ ; \quad |L,011\rangle \Rightarrow \frac{1}{\sqrt{2}} |L\rangle\label{fall apart L}\ , \\
|H,000\rangle &\Rightarrow& \frac{1}{2\sqrt{3}}|H\rangle\ ; \quad |H,011\rangle \Rightarrow \frac{1}{\sqrt{2}} |H\rangle\ .
\label{fall apart H}
\end{eqnarray}
So one can read off concrete fall-apart modes when $|L\rangle$, $|H\rangle$ are replaced by Eqs.~(\ref{modes_L}),(\ref{modes_H}).

Finally, we can find the fall-apart modes of the physical states by two steps.
First insert Eqs.~(\ref{fall apart L}), (\ref{fall apart H}) in Eqs.~(\ref{psi2spin000}), (\ref{psi2spin011}), (\ref{psi1spin000}), (\ref{psi1spin011}),
to get the fall-apart modes for $|\psi_2, 000\rangle$, $|\psi_2, 011\rangle$, $|\psi_1, 000\rangle$, $|\psi_1, 011\rangle$.
Then the resulting modes are substituted into Eqs.~(\ref{mixing2}),(\ref{mixing1}) to find the fall-apart modes
for $f_0(980)$, $f_0(1500)$, $f_0(500)$, $f_0(1370)$.
Since the technical steps are straightforward, here we simply give the final expressions for the fall-apart modes of the isoscalar resonances,
\begin{widetext}
\begin{eqnarray}
&&|f_0(500)\rangle:\Big\{\frac{1}{3}\left[(a+b\sqrt{2})\eta_1\eta_1+(a\sqrt{2}-b)\eta_1\eta_8
+(a/2-b\sqrt{2})\eta_8\eta_8\right]
-\frac{a}{2}\bm{\pi}\cdot\bm{\pi}-\frac{b}{\sqrt{2}}\overline{K} K
\Big\}\left (\frac{\beta_1}{2\sqrt{3}}+\frac{\alpha_1}{\sqrt{2}}\right )\label{500mode}\ ,\\
&&|f_0(1370)\rangle:\Big\{~~~~~~~~-\text{The same part appears here as the equation just above}-~~~~~~~~~~
\Big\}\left (-\frac{\alpha_1}{2\sqrt{3}}+\frac{\beta_1}{\sqrt{2}}\right )\label{1370mode}\ ,\\
&&|f_0(980)\rangle:\Big\{\frac{1}{3}\left[(-b+a\sqrt{2})\eta_1\eta_1 - (b\sqrt{2}+a)\eta_1\eta_8
-(b/2+a\sqrt{2})\eta_8\eta_8\right]
+\frac{b}{2}\bm{\pi}\cdot\bm{\pi}-\frac{a}{\sqrt{2}}\overline{K} K
\Big\}\left (\frac{\beta_2}{2\sqrt{3}}+\frac{\alpha_2}{\sqrt{2}}\right )\label{980mode}\ ,\\
&&|f_0(1500)\rangle:\Big\{~~~~~~~~~-\text{The same part appears here as the equation just above}-~~~~~~~~~
\Big\}\left (-\frac{\alpha_2}{2\sqrt{3}}+\frac{\beta_2}{\sqrt{2}}\right )\label{1500mode}\ .
\end{eqnarray}
\end{widetext}
These expressions look messy but they are simply related to each other by exchanging the mixing parameters involved.
For example, Eq.~(\ref{980mode}) can be obtained from Eq.~(\ref{500mode}) by $a\rightarrow -b, b\rightarrow a$,
$\alpha_1\rightarrow \alpha_2$, $\beta_1\rightarrow \beta_2$.
And Eq.~(\ref{1370mode}) can be obtained from Eq.~(\ref{500mode}) by $\alpha_1\rightarrow \beta_1, \beta_1 \rightarrow -\alpha_1$.
These simple replacements are just the consequences
of Eqs.(\ref{psi2spin000}),(\ref{psi2spin011}),(\ref{mixing2}),(\ref{psi1spin000}),(\ref{psi1spin011}),(\ref{mixing1}).
We identify the coefficient of each pseudoscalar mode in Eqs.~(\ref{500mode}), (\ref{1370mode}), (\ref{980mode}), (\ref{1500mode})
as the relative coupling strength of the corresponding resonance to that fall-apart mode.
Using the parameters given in Table~\ref{parameters} depending on the three cases, we determine
all the coupling strengths of possible fall-apart modes.
Their numeric values, up to an overall constant, are given in Table~\ref{fall apart modes}.

\begin{widetext}

\begin{table}
\centering
\begin{tabular}{c|c|c|c|c|c|c||c|c|c|c|c|c}  \hline\hline
\multirow{2}{*}{Channel} & \multicolumn{3}{c|}{$f_0(500)$} & \multicolumn{3}{c||}{$f_0(1370)$} & \multicolumn{3}{c|}{$f_0(980)$} & \multicolumn{3}{c}{$f_0(1500)$}\\
                         \cline{2-13}
                         &  SSC    & IMC &  RCF           &  SSC    & IMC &  RCF                &  SSC    & IMC &  RCF           &  SSC    & IMC &  RCF\\
\hline
$\pi^0\pi^0$             & -0.303 & -0.372 & -0.331    & -0.072 & -0.088 & -0.079         & -0.215 & 0.000 & -0.169      & -0.050 & 0.000 & -0.039\\[0.3mm]
$\pi^+\pi^-$             & -0.607 & -0.743 & -0.662    & -0.144 & -0.177 & -0.157         & -0.429 & 0.000 & -0.338     & -0.101 & 0.000 & -0.079\\[0.3mm]
$\overline{K}^0K^0$      & 0.303  & 0.000 & 0.239      & 0.072  & 0.000 & 0.057           & -0.429 & -0.526 & -0.469    & -0.101 & -0.122 & -0.109\\[0.3mm]
$K^+K^-$                 & 0.303  & 0.000 & 0.239      & 0.072  & 0.000 & 0.057           & -0.429 & -0.526 & -0.469    & -0.101 & -0.122 & -0.109\\[0.3mm]
\hline
$ \eta_8 \eta_8$         & 0.303  & 0.124 & 0.269      & 0.072  & 0.029 & 0.064           & -0.215 & -0.351 & -0.256    & -0.050 & -0.081 & -0.060\\[0.3mm]
$ \eta_1 \eta_8$         & 0.429  & 0.350 & 0.425      & 0.101  & 0.083 & 0.101           & 0.000  & -0.248 & -0.062    & 0.000  & -0.057 & -0.014\\[0.3mm]
$ \eta_1 \eta_1$         & 0.000  & 0.248 & 0.062      & 0.000  & 0.059 & 0.015           & 0.429  & 0.351 & 0.425      & 0.101  &  0.081 & 0.099\\[0.3mm]
\hline\hline
\end{tabular}
\caption{The relative coupling strengths of fall-apart modes for the isoscalar resonances with $J=0$,
calculated in the three different cases, SSC, IMC, RCF. For the actual couplings, an unknown overall factor must be multiplied.
We show the mode strengths of the lighter pair, $f_0(500)$, $f_0(1370)$ in the left panel and those of the heavy pair in the right panel.
As far as the nonzero couplings are concerned, we notice that, because of the spin configuration mixing,
the couplings of $f_0(500)$ and $f_0(980)$ are strongly enhanced compared to their counterparts in heavy nonets, $f_0(1370)$, $f_0(1500)$.
}
\label{fall apart modes}
\end{table}
\end{widetext}

As for the SSC result, we see that the SU(3)$_f$ relations are satisfied among most couplings considered.
For example, $f_0(500)$, being an isoscalar member of the octet,
satisfies the SU(3)$_f$ relation for its couplings with the octet members of the pseudoscalar meson,
$f_0(500) \pi^0\pi^0= f_0(500) K^+K^- = f_0(500) \eta_8 \eta_8=\frac{1}{2} f_0(500) \pi^+ \pi^-$, etc.
The similar relations can be seen from the $f_0(1370)$ couplings.
In addition, there are various vanishing modes that can be understood from the SU(3)$_f$ symmetry.
The $f_0(500)\eta_1\eta_1$, $f_0(1370)\eta_1\eta_1$ modes are zero because the octet members, $f_0(500)$, $f_0(1370)$,
are decoupled from the flavor singlet member, $\eta_1$.  Similarly, $f_0(980)\eta_1\eta_8$, $f_0(1500)\eta_1\eta_8$ are zero
because the flavor singlet members, $f_0(980)$, $f_0(1500)$, are decoupled from the singlet $\eta_1$ and the octet member $\eta_8$.
Due to the kinematical constraint, most of these vanishing modes can not be checked from experiments
except the mode, $f_0(1500)\eta_1\eta_8$.
But the zero coupling of $f_0(1500)\eta_1\eta_8$ is not consistent with the experimental fact.  In PDG, the $f_0(1500)$
has the decay mode of $f_0(1500)\rightarrow \eta_1\eta_8$ with a certain amount of branching ratio.
These results in SSC can be modified according to the generalized OZI rule that introduces the flavor
mixing between the octet and singlet.
Certainly in IMC, the $f_0(1500)\eta_1\eta_8$ coupling in Table~\ref{fall apart modes} is not zero,
and therefore the consistency with PDG can be recovered.

But the IMC results introduce other inconsistency with the phenomenology.
In particular, the couplings listed in Table~\ref{fall apart modes} show other vanishing modes in the IMC results.
For $f_0(500)$, $f_0(1370)$, their couplings to $\overline{K}^0 K^0$, $K^+ K^-$ are
zero. For $f_0(980)$, $f_0(1500)$, the couplings to $\pi^0\pi^0$, $\pi^+\pi^-$ are zero.
These can be easily understood because, in the IMC, strange quarks are completely decoupled from up and down quarks.
So $f_0(500)$ and $f_0(1370)$ do not fall apart into kaons because they are composed only by up and down quarks.
The other pair, $f_0(980)$ and $f_0(1500)$, do not fall apart into two pions because their flavor structure
is something like $~\sim [ds][\bar{d}\bar{s}] + [su][\bar{s}\bar{u}]$.
However, these vanishing modes are not consistent with the current experimental observations.
In PDG, the decay mode of $f_0(1370)\rightarrow \overline{K}K$ is listed.
Also the $\pi\pi$ mode is reported to be dominant in the decays of $f_0(980)$ and $f_0(1500)$.
In this sense, the IMC results are not satisfactory.

The RCF results seem to remove all these inconsistencies especially in comparison with the experimental decay modes.
Except for the modes that are not accessible kinematically,
all the nonzero modes presented here can be found in PDG also.
For $f_0(500)$, only the $\pi \pi$ mode is allowed kinematically and this is the dominant mode also in PDG.
For $f_0(1370)$, the decay channels, $\pi \pi$, $K\bar{K}$, $\eta_8 \eta_8$, are open kinematically and they have
been seen in experiments.  A similar consistency can be seen in the decay modes of $f_0(980)$, $f_0(1500)$.
In this sense, our tetraquark mixing framework in RCF is promising
as a realizable picture for those isoscalar resonances in this spin-0 channel.

The most striking feature of Table~\ref{fall apart modes} is that the couplings are strongly enhanced for $f_0(500)$, $f_0(980)$
while the corresponding couplings are suppressed for $f_0(1370)$, $f_0(1500)$.
As an example, the coupling for $f_0(500)\rightarrow \pi^0 \pi^0$ is -0.331 while $f_0(1370)\rightarrow \pi^0 \pi^0$ is -0.079,
about a factor 4 smaller.
This result originates from the mixing formulas, Eq.~(\ref{mixing2}) for $f_0(980)$, $f_0(1500)$,
and Eq.~(\ref{mixing1}) for $f_0(500)$, $f_0(1370)$.
For instance, in Eq.~(\ref{mixing2}), one can see that, due to the relative sign difference, the two spin configurations
cancel in making $|f_0(1500)\rangle$ while they add up in making $|f_0(980)\rangle$.  These cancelation and addition
still persist even when the two spin configurations simply fall apart into two mesons,
which then yields the enhancement and suppression of the couplings.
In principle, these results can be tested by experiments through the measurement of partial decay widths.
In fact, this type of phenomena tested in the isovector channel is found to be consistent with the
experimental data~\cite{Kim:2017yur}.
Unfortunately, at present, this comparison is not possible for the isoscalar channel due to the limited experimental information.
For the isoscalar resonances, PDG shows those decay modes only without providing specific numbers
for most partial decay widths.
Only the resonance $f_0(1500)$ is the one that have the measured branching ratios but it is not enough to test the
main result of our model, the enhancement and suppression due to the mixing framework.
Thus, our interesting results can not be tested at the present situation.

\subsection{For spin-1 tetraquarks}

As a further supporting evidence for the tetraquark mixing framework,
we have proposed in Sec.~\ref{sec:spin12} the additional tetraquarks to be found in the spin-1 channel.
For the isoscalar members, candidates are chosen to be $h_1 (1170)$, $h_1 (1380)$.
In Sec.~\ref{sec:hf_spin12}, we reported that
their hyperfine mass splitting from the corresponding members in spin-0 tetraquarks has a similar trend with the experimental
mass splitting even though the agreement is not precise.
In this subsection, we test tetraquark structure of the spin-1 candidates further
by investigating their fall-apart decay modes
and comparing them with the experimental decay modes from $h_1 (1170)$, $h_1 (1380)$.

The spin configuration of the $J=1$ tetraquarks is $|J J_{12} J_{34}\rangle = |111\rangle$ in the diquark ($J_{12}$)
and antidiquark ($J_{34}$) spin bases. For the fall-apart modes, we need to
rearrange this spin state in terms of the (13), (24) pairs.
For our demonstration, we take the state with the maximal spin projection, $J=1, M=1$,
among three spin states of $|111\rangle$, i.e., $|JM\rangle=|11\rangle, |10\rangle, |1-1\rangle$.  But our discussion
below must be irrespective to this choice.
We can readily write the state $|11\rangle$
with respect to the spin states of the quark-antiquark pairs, $|J_{13} M_{13}\rangle$, $|J_{24} M_{24}\rangle$, as
\begin{eqnarray}
|11 \rangle &=& \frac{1}{\sqrt{2}} \left[ |11\rangle_{13} |00\rangle_{24} +  |00\rangle_{13} |11\rangle_{24} \right ]  \label{spinbases}\ .
\end{eqnarray}
So one can see that the fall-apart modes are divided into two categories,
the first type composed by a vector meson from the (13) pair and a pseudoscalar meson from the (24) pair,
and the second type composed by the other way around.

Using the flavor structures of the ideal mixing states, $|L,111\rangle$, $|H,111\rangle$,
and folding them into Eqs.~(\ref{psi11}), (\ref{psi21}), we readily evaluate the
fall-apart modes of $|h_1(1170) \rangle$, $|h_1(1380) \rangle$,
\begin{eqnarray}
|h_1(1170) \rangle &:& \left (\frac{a}{\sqrt{3}}+\frac{b}{\sqrt{6}} \right ) \omega\eta_1
+ \left (\frac{a}{\sqrt{6}}-\frac{b}{\sqrt{3}} \right ) \omega\eta_8 \nonumber\\
&&+ \frac{b}{\sqrt{3}}\left (\phi \eta_1 + \frac{1}{\sqrt{2}}\phi \eta_8\right ) -\frac{a}{\sqrt{2}}\bm{\rho}\cdot\bm{\pi} \nonumber \\
&&-\frac{b}{\sqrt{2}}\left (\overline{K}^* K + \overline{K} K^*\right )\label{1170modes}\ ,\\
|h_1(1380) \rangle &:& \left ( \text{replacing}~a \rightarrow -b, b\rightarrow a \right )\label{1380modes}\ .
\end{eqnarray}
Here the isodoublets for the $K^*$ are defined similarly as Eq.~(\ref{isodoublets}).
In deriving these, the ideal mixing is assumed for the $\phi$ and $\omega$ so that
$\phi=s\bar{s}$, $\omega=\frac{1}{\sqrt{2}}[u\bar{u}+d\bar{d}]$.
The color factor, obtained from the formation of the color singlets from the (13) and (24) pairs, is common for
all the terms here so it can be absorbed into the overall factor.

Again using the mixing parameters given in Table~\ref{parameters}, we calculate
the relative coupling strengths for all the fall-apart modes and list them
in Table~\ref{spin1 fall apart modes}.
For $h_1(1170)$, most channels are not allowed kinematically. The only channel allowed
is $h_1(1170)\rightarrow \rho \pi$ and, in fact, it is supported by PDG that shows $h_1(1170)\rightarrow \rho\pi$ as a sole
measured decay mode.
So our fall-apart mode of $h_1(1170)$ is not inconsistent with experimental situation.

For $h_1(1380)$, we have the three modes, $\rho \pi$, $\omega \eta_8$, $K^* K$, that are allowed kinematically.
The third mode is barely allowed as the $h_1(1380)$ mass is close to the $K^* K$ threshold.
The fall-apart decay modes based on our calculation give nonzero couplings for most channels.
Only the $\rho \pi$ mode in IMC gives its coupling zero which can be understood from its
flavor structure in IMC, $\sim [ds][\bar{d}\bar{s}] + [su][\bar{s}\bar{u}]$.
These fall-apart modes do not agree with the current experimental decay modes for $h_1(1380)$.
According to PDG, $h_1(1380)$ has one decay mode, $K^* K$, only.
So we have the inconsistency problem for the decays of $h_1(1380)$ when it is viewed in
the tetraquark picture.

To understand this inconsistency, one can contemplate various possibilities. One possibility is that
$h_1(1380)$ is not the anticipated spin-1 tetraquark.  As we discussed briefly earlier,
its isospin is not determined according to PDG so this may not be the isoscalar resonance.
The other candidate in PDG, $h_1(1595)$, was excluded
as a candidate for the spin-1 tetraquark because its mass is too heavy.
Thus, if $h_1(1380)$ is not the right candidate, one can expect other resonance to
be discovered in this channel in future. To get the better agreement in mass splitting, the new resonance hopefully needs to have the mass around 1340 MeV.
The other possibility is that the candidate is hidden in the two-meson continuum meaning that the spin-1 tetraquark is too broad to
appear as a resonance structure in PDG.
Another possibility, which is the best for us, might be that the missing modes from $h_1(1380)$ appear in future experiments.
Anyway, according to our analysis, we have some hints for the existence of the spin-1 tetraquark,
like the mass splitting and some modes for their decays, but they are not conclusive at the present situation.

\begin{table}
\centering
\begin{tabular}{c|c|c|c|c|c|c}  \hline\hline
\multirow{2}{*}{Channel} & \multicolumn{3}{c|}{$h_1(1170)$} & \multicolumn{3}{c}{$h_1(1380)$} \\
                         \cline{2-7}
                         &  SSC    & IMC &  RCF           &  SSC    & IMC &  RCF                \\
\hline
$\rho^0\pi^0$             & -0.577 & -0.707 & -0.630    & -0.408 & 0.000 & -0.321         \\[0.3mm]
$K^{*-}K^+$               & 0.289  & 0.000  & 0.227     & -0.408  & -0.500 & -0.445           \\[0.3mm]
\hline
$ \omega \eta_1$          & 0.236  & 0.577 & 0.329     & 0.667  & 0.408 & 0.626           \\[0.3mm]
$ \omega \eta_8$          & 0.667  & 0.408 & 0.626     & -0.236 & -0.577 & -0.329           \\[0.3mm]
$ \phi \eta_1$            & -0.333 & 0.000 & -0.262     & 0.471  & 0.577 & 0.514           \\[0.3mm]
$ \phi \eta_8$            & -0.236 & 0.000 & -0.185     & 0.333  & 0.408 & 0.364           \\[0.3mm]
\hline\hline
\end{tabular}
\caption{The relative coupling strengths of fall-apart modes for the isoscalar tetraquarks with $J=1$,
calculated in the three different cases, SSC, IMC, RCF. It should be understood that, for the actual couplings,
an unknown overall factor must be multiplied.
The isospin multiplet pairs have the same couplings. That is, the $\rho^0\pi^0$ coupling is the same
as the $\rho^+\pi^-$, $\rho^-\pi^+$ couplings.
Also the $K^{*-}K^+$ coupling is the same as the couplings of $\overline{K}^{*0}K^0$, $K^{*0}\overline{K}^0$, $K^{*+}K^-$.
}
\label{spin1 fall apart modes}
\end{table}

\section{Summary}
\label{sec:summary}

In this work, we have performed an intensive investigation on the tetraqaurk possibility for light mesons
especially in the isoscalar channel.
First, based on an observation that there are two nonets in meson spectra in the $J^P=0^+$ channel,
we have constructed two tetraquarks either by the spin-0 diquark or by the spin-1 diquark
in the diquark-antidiquark form.
The two tetraquarks differ by the spin and color configurations,
one type as $|000\rangle$, $|\bm{1}_c,\bar{\bm{3}}_c,\bm{3}_c\rangle$, and the other type as
$|011\rangle$, $|\bm{1}_c,\bm{6}_c,\bar{\bm{6}}_c\rangle$.
The most important aspect is that the two configurations mix strongly through the color-spin interactions which,
under the diagonalization, can generate the physical resonances that can be identified as the two nonets in PDG.
Specifically, we have applied the configuration mixing between $f_0(500)$, $f_0(980)$ in the light nonet
and $f_0(1370)$, $f_0(1500)$ in the heavy nonet.

One complication in this isoscalar channel is how to implement
the additional flavor mixing normally known as the OZI rule.  To take into account this flavor mixing,
we have considered three different cases, SU(3)$_f$ symmetric case (SSC), ideal mixing case(IMC),
the realistic case with fitting (RCF).

Our results for $f_0(980)$, $f_0(1500)$ show that there is a huge separation in hyperfine
masses which is consistent with the mass splitting between $f_0(980)$, $f_0(1500)$.
For $f_0(500)$, $f_0(1370)$, since the experimental masses contain large error bars,
we can not make a precise comparison but the hyperfine mass splitting is still huge, qualitatively
agreeing with the mass splitting. These results are found to be insensitive to how the flavor mixing is implemented.
We have found that the huge separation in masses are mainly driven by the spin configuration mixing.
This could be a strong indication that the tetraquark mixing framework is realized by the two nonets in light mesons.

Our tetraquark wave functions in the $J^P=0^+$ channel can be used to study
the decay patterns of the isoscalar resonances through
their fall-apart decay modes. Considering the experimental accessibility for comparison,
we have focussed on the decays into two pseudoscalar mesons and
presented possible modes calculated in the three different cases of the flavor mixing.
The most striking feature is that the coupling strengths are enhanced for
the resonances belonging to the light nonet while they are suppressed for the resonances in the heavy nonet.
Due to the scarcely known branching ratios of the resonances, this interesting consequence can not
be compared with the experimental data.
We anticipate that this result must be tested in future experiments.
The fall-apart modes, calculated in SSC, IMC, are found to have some vanishing modes which
do not agree with the experimental decay modes.  But the modes found from RCF
are consistent with the experimental modes.

To support the tetraquark mixing framework, it is necessary that spin-1,2 tetraquarks exist also.
For the spin-2 isoscalar candidates, there are various resonances that one can choose from PDG but the selection
involves some arbitrariness. For the spin-1 isoscalar candidates, we take $h_1(1170)$, $h_1(1380)$
and test whether they are consistent with the tetraquark picture by calculating the mass splitting
and the fall-apart modes.  The mass splitting has a rough agreement with the hyperfine mass splitting,
thus supporting the tetraquark picture.  The fall-apart modes for $h_1(1170)$ also is not inconsistent with
the experimental situation.  But there are disagreements for the decay modes of $h_1(1380)$ and
we have discussed possible resolutions.

\acknowledgments

\newblock
The work of H.Kim was supported by Basic Science Research Program through the National Research Foundation of Korea(NRF)
funded by the Ministry of Education(Grant No. 2015R1D1A1A01059529).
The work of K.S.Kim was supported by the National Research Foundation of Korea (Grant No. 2015R1A2A2A01004727).
The work of M.Oka is supported by the Grant-in-Aid for Scientific Research (No.~25247036) from Japan Society for the Promotion of Science (JSPS).

\appendix*
\section{$C$-parity of the tetraquarks}

Here we determine the $C$-parity of the $I_z=0$ members in the tetraquark nonet in $J=0,1,2$ channels.
There are three members with $I_z=0$ in the tetraquark nonet, two isoscalars and one isovector.
Their flavor structure has a common feature that the diquark and antidiquark parts are connected by the charge conjugation.
[See the flavor structure of the two isoscalars given in Eqs.~(\ref{singlet}), (\ref{octet}).]
So, to determine $C$-parity, it is enough to consider one specific flavor combination and we take the
part, $[su][\bar{s}\bar{u}]$, for an illustration purpose.

For the $J=0$ tetraquark, we have two spin configurations, $|J,J_{12},J_{34}\rangle = |000\rangle, |011\rangle$.
If we denote the states by its total spin and its
projection, $J$ and $M$, both are in $|J,M\rangle=|0,0\rangle$, and, to distinguish two spin configurations,
we label them as $|000\rangle=|0,0\rangle_a$, $|011\rangle=|0,0\rangle_b$.
Because they are composed by the diquark and antidiquark,
the spin states, $|0,0\rangle_a, |0,0\rangle_b$, can be expressed by the spins and their projections of diquark and antidiquark,
$|J_{12},M_{12}\rangle_{[su]} |J_{34},M_{34}\rangle_{[\bar{s}\bar{u}]}$. Namely, we have the followings,
\begin{eqnarray}
|0,0\rangle_a &=& |0,0\rangle_{[su]} |0,0\rangle_{[\bar{s}\bar{u}]}  \ ,\\
|0,0\rangle_b &=& \frac{1}{\sqrt{3}} \Big \{ |1,1\rangle_{[su]} |1,-1\rangle_{[\bar{s}\bar{u}]} - |1,0\rangle_{[su]} |1,0\rangle_{[\bar{s}\bar{u}]} \nonumber \\
& &+ |1,-1\rangle_{[su]} |1,1\rangle_{[\bar{s}\bar{u}]}] \Big \} \label{00b}\ ,
\end{eqnarray}
with the appropriate Clebsch-Gordan (CG) coefficients.
Under $C$, the diquark and antidiquark will be interchanged, $[su]\leftrightarrow [\bar{s}\bar{u}]$ and we have
for $|0,0\rangle_a$,
\begin{eqnarray}
C|0,0\rangle_a &=& |0,0\rangle_{[\bar{s}\bar{u}]} |0,0\rangle_{[su]} = |0,0\rangle_a \ .
\end{eqnarray}
So the $C$-parity of $|0,0\rangle_a$ is even.
For $|0,0\rangle_b$, under the charge conjugation, it becomes
\begin{eqnarray}
C|0,0\rangle_b &=& \frac{1}{\sqrt{3}} \Big \{ |1,1\rangle_{[\bar{s}\bar{u}]} |1,-1\rangle_{[su]} - |1,0\rangle_{[\bar{s}\bar{u}]} |1,0\rangle_{[su]} \nonumber \\
& &+ |1,-1\rangle_{[\bar{s}\bar{u}]} |1,1\rangle_{[su]}] \Big \} \ .
\end{eqnarray}
That is, the charge conjugation only switches the first and third terms in $|0,0\rangle_b$ which is the same
with the original state that we have started with.
Therefore, the $J=0$ tetraquark has $C=+$.

For the $J=1$ tetraquark,  the spin configuration is $|111\rangle$ and one can prove that $C|111\rangle=-|111\rangle$.
To do that, we take the state with $J=1$ and the spin projection $M=1$, $|J,M\rangle =|1,1\rangle$.
Again, we can write this state in terms of the spins and their projections of diquark and antidiquark,
$|J_{12},M_{12}\rangle_{[su]} |J_{34},M_{34}\rangle_{[\bar{s}\bar{u}]}$ with accompanying CG coefficients as
\begin{eqnarray}
|1,1\rangle=\frac{1}{\sqrt{2}}\Big\{|1,1\rangle_{[su]}|1,0\rangle_{[\bar{s}\bar{u}]}
- |1,0\rangle_{[su]}|1,1\rangle_{[\bar{s}\bar{u}]}\Big\}\nonumber\ .
\end{eqnarray}
Under $C$, it becomes
\begin{eqnarray}
C|1,1\rangle &=&\frac{1}{\sqrt{2}}\left\{|1,1\rangle_{[\bar{s}\bar{u}]}|1,0\rangle_{[su]}- |1,0\rangle_{[\bar{s}\bar{u}]}|1,1\rangle_{[su]}\right\}
\nonumber\ .
\end{eqnarray}
Rearranging the diquark part in the front and the antidiquark part in the back leads to the state whose
overall sign is opposite to the original state. This means that the $C$-parity of $|1,1\rangle$ is odd, $C=-$.

For the $J=2$ tetraquark, the spin configuration is $|211\rangle$. There are five spin states differed by its spin projection.
Since all the five states have the same $C$-parity,
we can consider one of them, for example, the state with maximal spin projection, $|J,M\rangle =|2,2\rangle$.
This state is expressed as
\begin{eqnarray}
|2,2\rangle_a &=& |1,1\rangle_{[su]} |1,1\rangle_{[\bar{s}\bar{u}]}  \ ,
\end{eqnarray}
in terms of diquark and antiquark spins and their projections.  Then, similarly as above,
it is easy to see that the $C$-parity of $|2,2\rangle$ is even.

\end{document}